\documentclass{article}
\usepackage{proceed2e}
\usepackage{url}
\usepackage{graphicx}
\usepackage[round]{natbib}
\usepackage{amsmath}
\usepackage{amssymb}
\usepackage{algorithm}
\usepackage{algorithmicx}
\usepackage{algpseudocode}

\usepackage{amsmath}
\usepackage{amsfonts}
\usepackage{bbm} 

\DeclareMathOperator*{\argmin}{argmin}

\newcommand{\bbE}{\mathbb{E}}

\newcommand{\bbone}{\mathbbm{1}}

\newcommand{\cH}{\mathcal{H}}

\newcommand{\cO}{\mathcal{O}}

\newcommand{\nn}{\nonumber}

\newcommand{\binomialCoefficient}[2]{\left(\begin{array}{c}  #1 \\ #2  \end{array}\right)}

\usepackage{amsthm} 

\newtheorem{theorem}{Theorem}[section]
\newtheorem{lemma}[theorem]{Lemma}

\title{Quantum Annealing for Clustering} 
 
\author{
{\bf Kenichi Kurihara} \\  
Google,\\  
Tokyo, Japan\\ 
\And 
{\bf Shu Tanaka} \\
Institute for Solid State Physics,\\
University of Tokyo\\
Chiba, Japan
\And 
{\bf Seiji Miyashita} \\
Dept. of Physics,\\
University of Tokyo, Tokyo, Japan\\
CREST, Saitama, Japan\\
} 
 
\begin{document} 
 
\maketitle 
 
\begin{abstract} 
\vspace{-2mm}
  This paper studies quantum annealing (QA) for clustering, which
  can be seen as an extension of simulated annealing (SA).  We
  derive a QA algorithm for clustering and propose an annealing
  schedule, which is crucial in practice.  Experiments show the
  proposed QA algorithm finds better clustering assignments than SA.
  Furthermore, QA is as easy as SA to implement.
\vspace{-2mm}
\end{abstract}

\section{Introduction} 
Clustering is one of the most popular methods in data mining.
Typically, clustering problems are formulated as optimization
problems, which are solved by algorithms, for example the EM algorithm
or convex relaxation.  However, clustering is typically NP-hard.  The
simulated annealing (SA) \citep{Kirkpatrick83} is a promising
candidate. \citet{Geman84} proved SA was able to find the global optimum
with a slow cooling schedule of temperature $T$.
Although their schedule is in practice too slow for clustering of
a large amount of data, it is well known that SA still finds a reasonably good
solution even with a faster schedule than what \citeauthor{Geman84} proposed.

\begin{figure*}[t!]
  \begin{center}
  \begin{minipage}[b]{0.34\linewidth}
\begin{center}
\includegraphics[width=.8\linewidth]{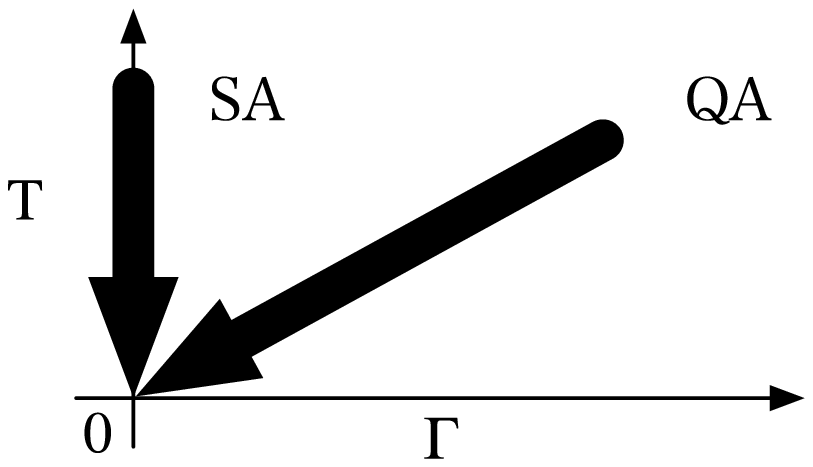}
\end{center}
\caption{Quantum annealing (QA) adds another dimension to 
  simulated annealing (SA) to control a model.  QA iteratively
  decreases $T$ and $\Gamma$ whereas SA decreases just $T$.}
\label{fig:pathes}
  \end{minipage}
  \hspace{.5cm}
  \begin{minipage}[b]{0.51\linewidth}
\begin{center}
\includegraphics[width=.99\linewidth]{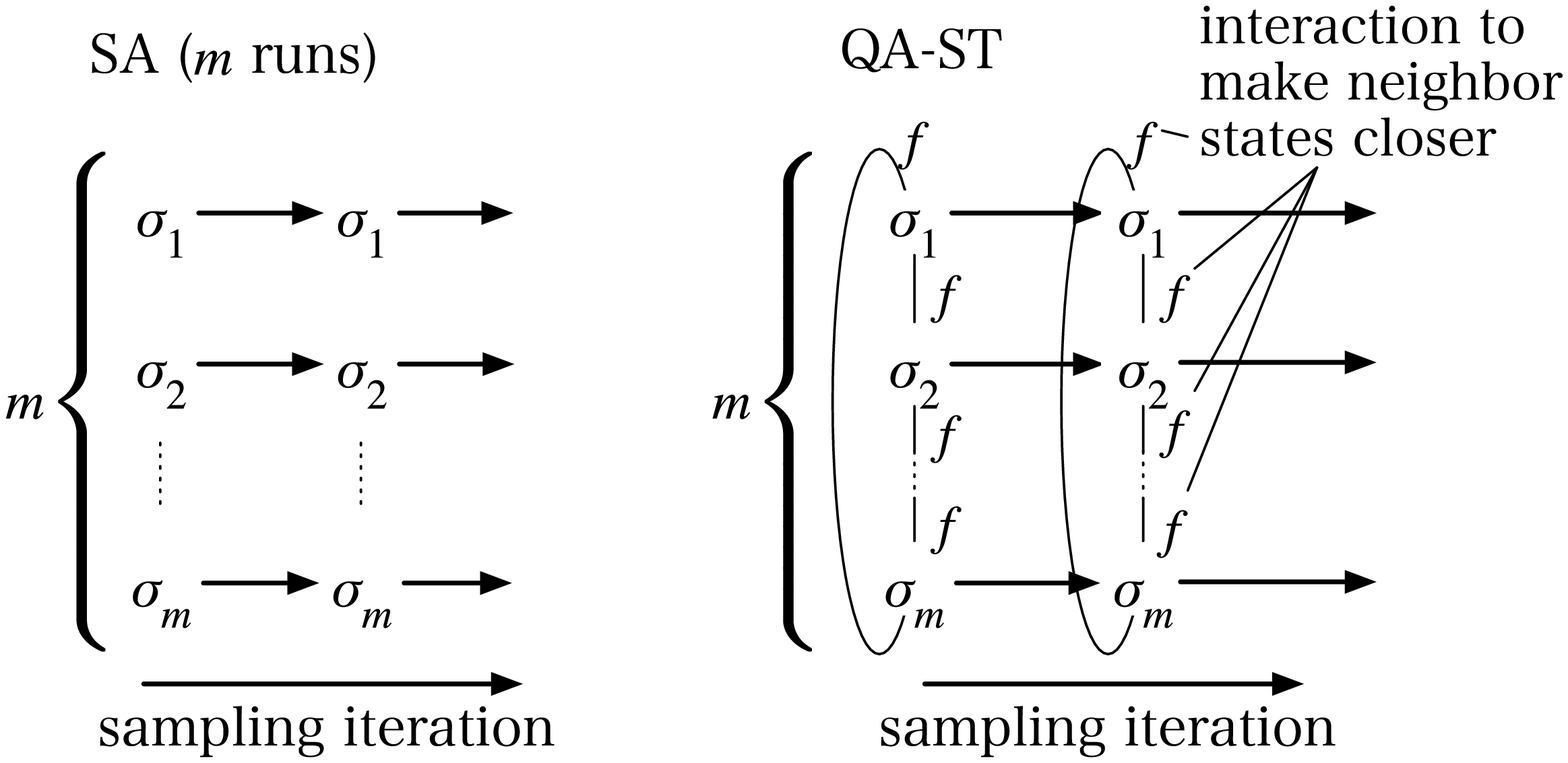}
\vspace{-6mm}
\end{center}
\vspace{-1mm}
\caption{Illustrative explanation of QA.  The left figure shows $m$
  independent SAs, and the right one is QA algorithm derived with the
  Suzuki-Trotter (ST) expansion.  $\sigma$ denotes a clustering
  assignment.}
\label{fig:sa_qa_update}
  \end{minipage}
\end{center}
\vspace{-3mm}
\end{figure*}

In statistical mechanics, quantum annealing (QA) has been proposed as
a novel alternative to SA \citep{Apolloni89,Kadowaki98,Santoro02,matsuda08}.  QA
adds another dimension, $\Gamma$, to SA for annealing, see
Fig.\ref{fig:pathes}.  Thus, it can be seen as an extension of SA.  QA
has succeeded in specific problems, e.g. the Ising model in
statistical mechanics, and it is still unclear that QA works better
than SA in general.  We do not actually think QA intuitively helps
clustering, but we apply QA to clustering just as procedure to derive
an algorithm.  A derived QA algorithm depends on the definition of
quantum effect $\cH_\text{q}$.  We propose quantum effect
$\cH_\text{q}$, which leads to a search strategy fit to clustering.
Our contribution is,
\vspace{-3mm}
\begin{enumerate}
  \setlength{\parskip}{-0mm} 
  \setlength{\itemsep}{-0mm} 
\item to propose a QA-based optimization algorithm for clustering, in
  particular
  \begin{enumerate}
    \item quantum effect $\cH_\text{q}$ for clustering
    \item and a good annealing schedule, which is crucial for
      applications,
  \end{enumerate}
\item and to experimentally show the proposed algorithm optimizes
  clustering assignments better than SA.
\end{enumerate}
\vspace{-3mm}
We also show the proposed algorithm is as easy as SA to
implement.



The algorithm we propose is a Markov chain Monte Carlo (MCMC) sampler,
which we call QA-ST sampler.  As we explain later, a naive QA sampler is
intractable even with MCMC.  Thus, we approximate QA by the
Suzuki-Trotter (ST) expansion \citep{Trotter59,Suzuki76} to derive a
tractable sampler, which is the QA-ST sampler.  QA-ST looks like
parallel $m$ SAs with interaction $f$ (see
Fig.\ref{fig:sa_qa_update}).  At the beginning of the annealing
process, QA-ST is almost the same as $m$ SAs.  Hence, QA-ST finds $m$
(local) optima independently.  As the annealing process continues,
interaction $f$ in Fig.\ref{fig:sa_qa_update} becomes stronger to move
$m$ states closer.  QA-ST at the end picks up the state with
the lowest energy in $m$ states as the final solution.

\begin{figure}[t!]
\vspace{-4mm}
\begin{center}
  \begin{tabular}{cc}
\multicolumn{2}{c}{
  \includegraphics[width=.95\linewidth]{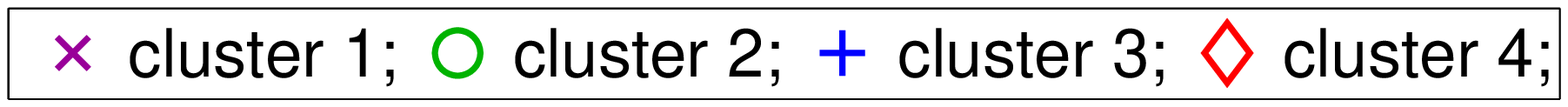}
}
\\
    $\sigma_1$ (local optimum)
    &
    $\sigma_2$ (local optimum)
    \\
\fbox{\includegraphics[width=.41\linewidth]{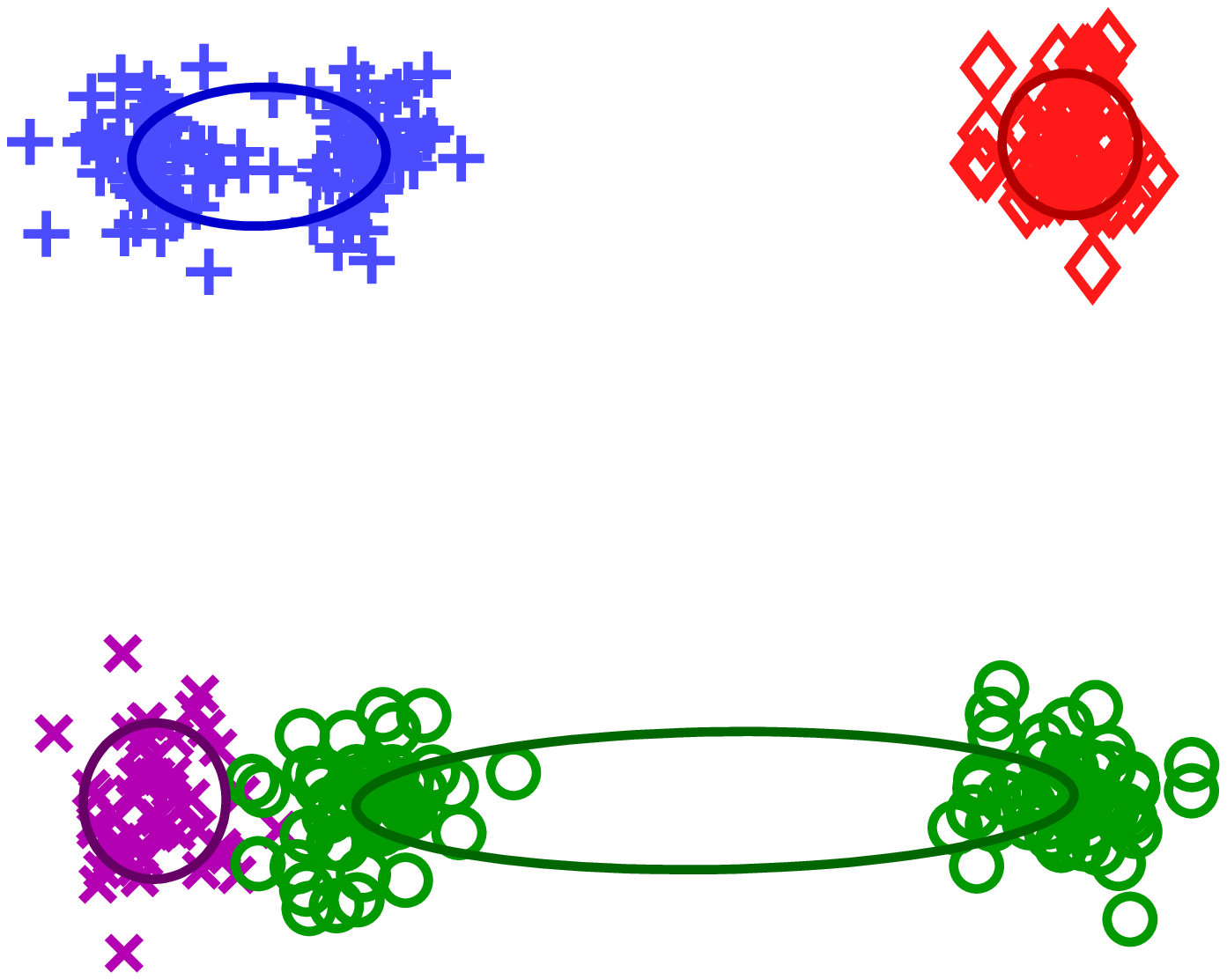}}
&
\fbox{\includegraphics[width=.41\linewidth]{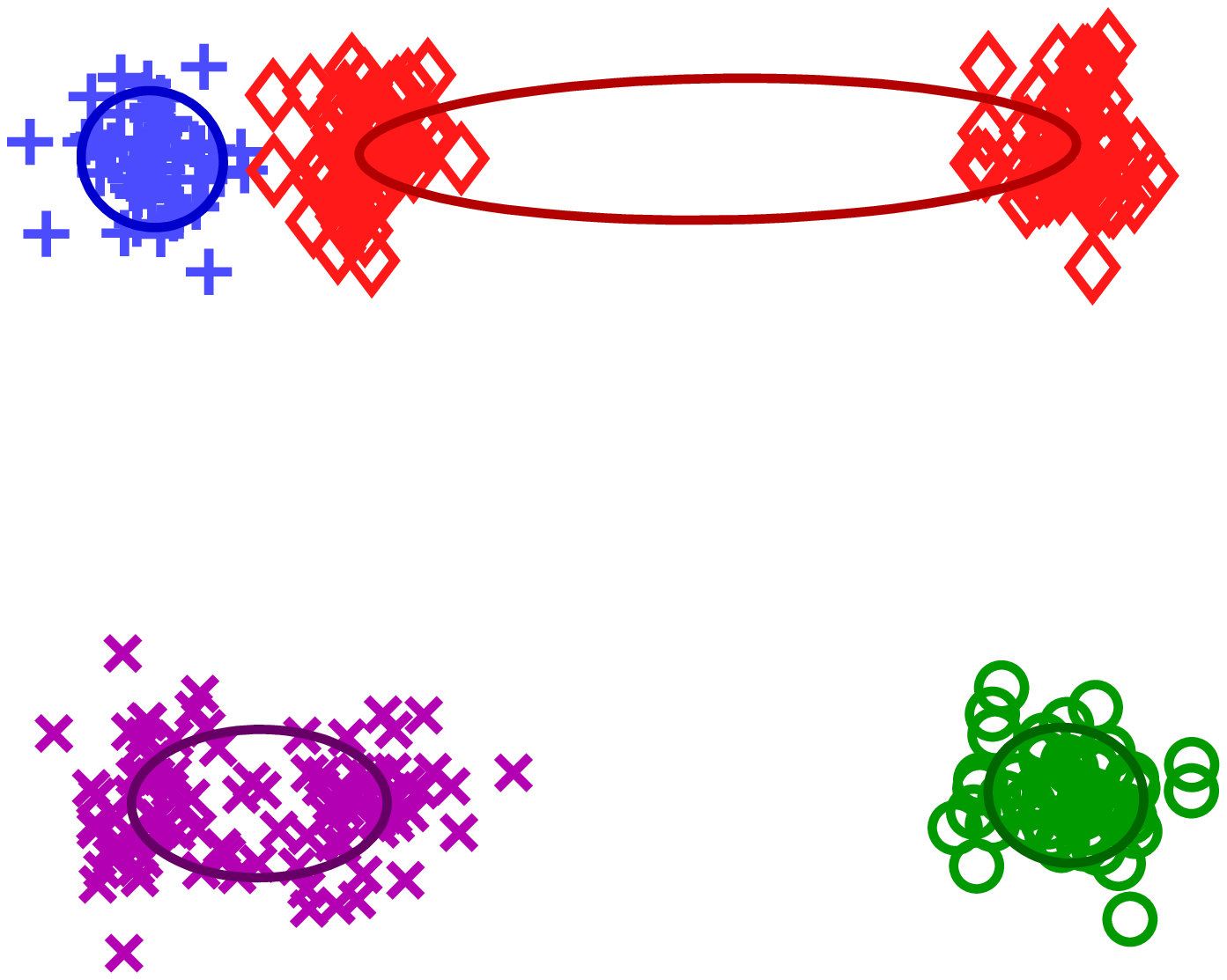}}
    \vspace{5mm}
\\
    $\sigma^*$ (global optimum)
    &
    \\
\fbox{\includegraphics[width=.41\linewidth]{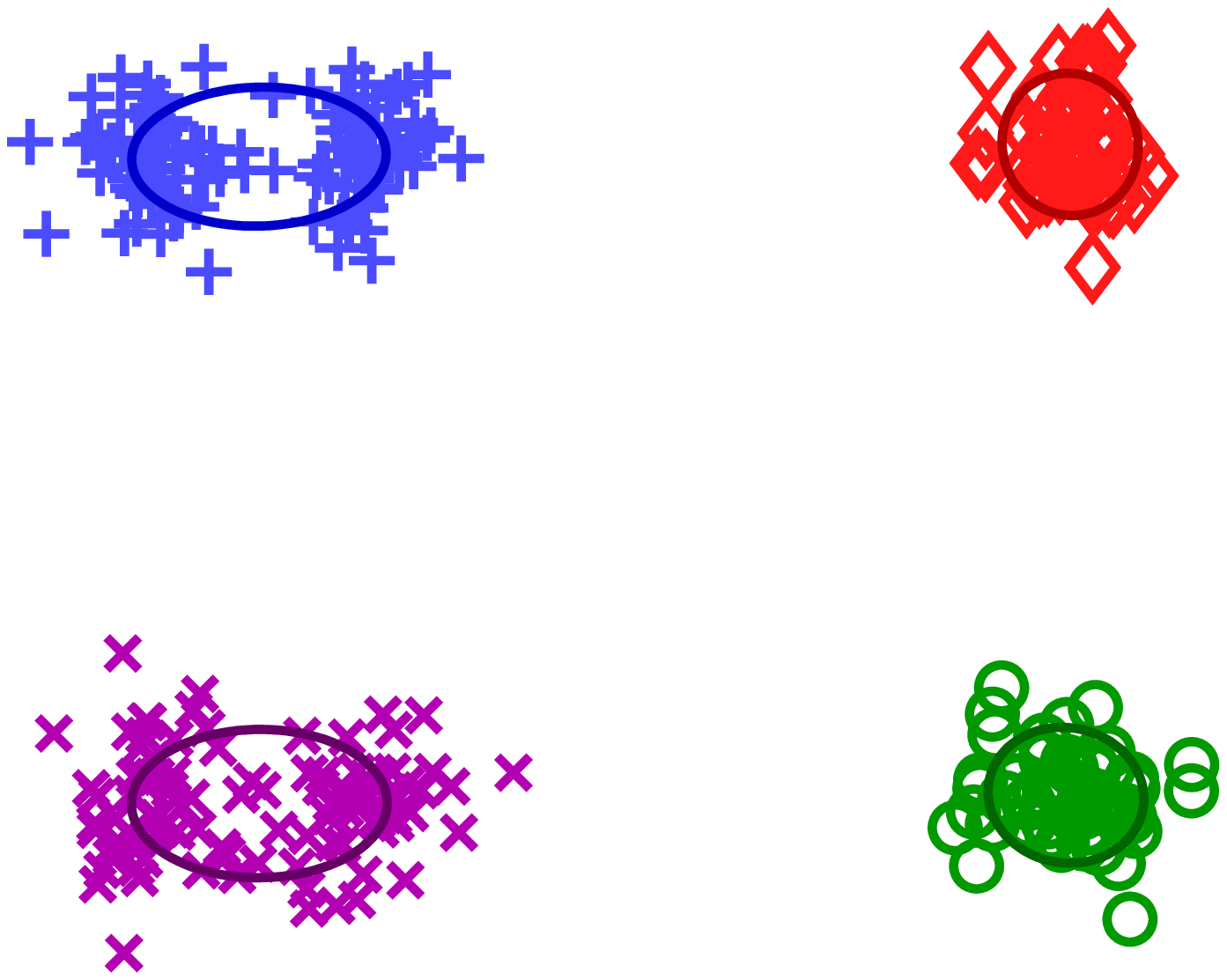}}
&
\hspace{-5mm}
\includegraphics[width=.41\linewidth]{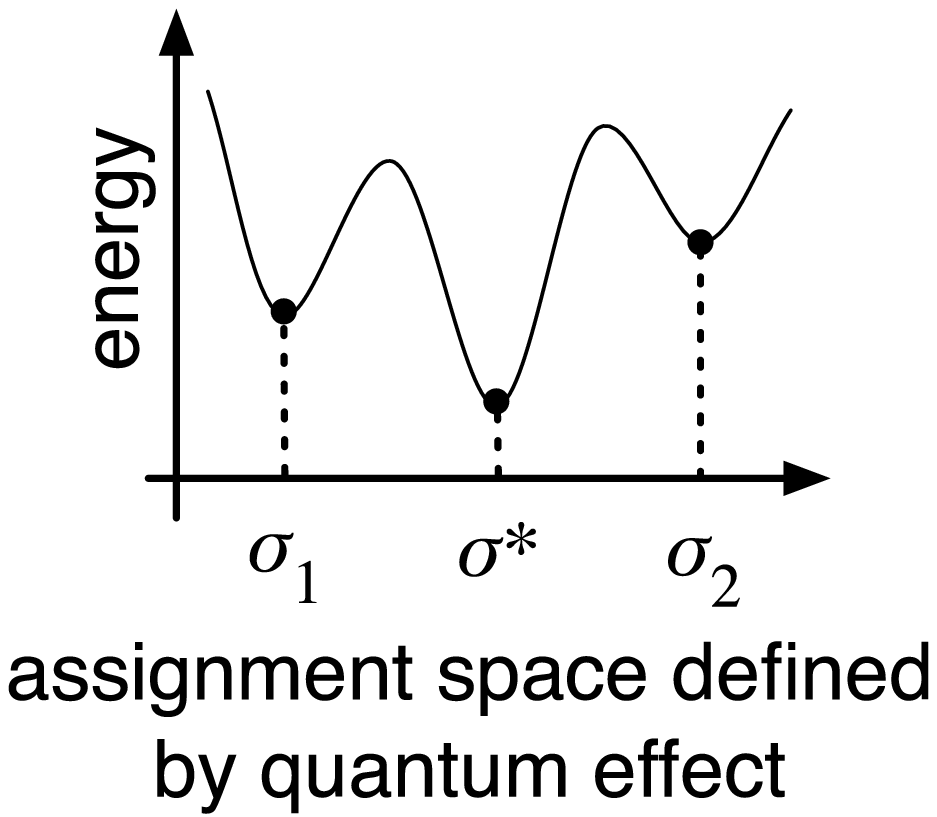}
  \end{tabular}
\end{center}
\vspace{-1mm}
\caption{Three clustering results by a mixture of four Gaussians
  (i.e. \#clusters=4).}
\label{fig:local_global}
\vspace{-4mm}
\end{figure}

QA-ST with the proposed quantum effect $\cH_\text{q}$ works well for
clustering.  Fig.\ref{fig:local_global} is an example where data
points are grouped into four clusters.  $\sigma_1$ and $\sigma_2$ are
locally optimal and $\sigma^*$ is globally optimal.  Suppose $m$ is
equal to two and $\sigma_1$ and $\sigma_2$ in
Fig.\ref{fig:sa_qa_update} correspond to $\sigma_1$ and $\sigma_2$ in
Fig.\ref{fig:local_global}.  Although $\sigma_1$ and $\sigma_2$ are
local optima, the interaction $f$ in Fig.\ref{fig:sa_qa_update} allows
$\sigma_1$ and $\sigma_2$ to search for a better clustering assignment
between $\sigma_1$ and $\sigma_2$.  Quantum effect $\cH_\text{q}$
defines the distance metric of clustering assignments.  In this case,
the proposed $\cH_\text{q}$ locates $\sigma^*$ between $\sigma_1$ and
$\sigma_2$.  Thus, the interaction $f$ gives good chance to go to
$\sigma^*$ because $f$ makes $\sigma_1$ and $\sigma_2$ closer (see
Fig.\ref{fig:sa_qa_update}).
The proposed algorithm actually finds
$\sigma^*$ from $\sigma_1$ and $\sigma_2$.  Fig.\ref{fig:local_global}
is just an example.  However, a similar situation often occurs in
clustering.  Clustering algorithms in most cases give ``almost''
globally optimal solutions like $\sigma_1$ and $\sigma_2$, where the
majority of data points are well-clustered, but some of them are not.
Thus, a better clustering assignment can be constructed by picking up
well-clustered data points from many sub-optimal clustering
assignments.  Note an assignment constructed in such a way is located
between the sub-optimal ones by the proposed quantum effect
$\cH_\text{q}$ so that QA-ST can find a better assignment between
sub-optimal ones.



\section{Preliminaries} \label{sec:preliminaries}
First of all, we introduce the notation used in this paper.  We assume we
have $n$ data points, and they are assigned to $k$ clusters.  The
assignment of the $i$-th data point is denoted by binary indicator
vector $\tilde{\sigma}_i$.  For example, when $k$ is equal to two,
we denote the $i$-th data point assigned to the first and the second cluster by
$\tilde{\sigma}_i = (1,~0)^T$ and $\tilde{\sigma}_i = (0,~1)^T$, respectively.
The assignment of all data points is also denoted by an indicator vector, $\sigma$, whose length
is $k^n$ because the number of available assignments is $k^n$.
$\sigma$ is constructed with $\{ \tilde{\sigma}_i \}_{i=1}^n$, 
$
  \sigma = \bigotimes_{i=1}^n \tilde{\sigma}_i
$,
where $\otimes$ is the Kronecker product, which is a special case of the tensor product
for matrices.  Let $A$ and $B$ be matrices where
$
A \!=\! \left(\begin{matrix}
  a_{11} \!& a_{12} \\
  a_{21} \!& a_{22}
\end{matrix}\right)
$.  Then, $A \otimes B \!=\! 
\left(\begin{matrix}
  a_{11}B \!& a_{12}B \\
  a_{21}B \!& a_{22}B
\end{matrix}\right)
$ (see \cite{Minka00} for example).
Only one element in $\sigma$ is one, and the others are
zero.
For example,
$\sigma = \tilde{\sigma}_1 \otimes \tilde{\sigma}_2 = (0,~1,~0,~0)^T$ 
when $k=2$, $n=2$, the first data point is assigned to the first
cluster ($\tilde{\sigma}_1 = (1,~0)^T$) and the second data point is
assigned to the second cluster ($\tilde{\sigma}_2 = (0,~1)^T$).
We also use $k$ by $n$ matrix $Y$ to denote the
assignment of all data, 
\begin{align}
Y(\sigma) = 
( \tilde{\sigma}_{1}, \tilde{\sigma}_{2}, ... , \tilde{\sigma}_{n} )
\label{eq:Y}
.
\end{align}
We do not store $\sigma$ in memory whose length is $k^n$, but we store
$Y$.  We use $\sigma$ only for the derivation of quantum
annealing.  The proposed QA algorithm is like parallel $m$ SAs.  We
denote the $j$-th SA of the parallel SA by $\sigma_j$.  The $i$-th
data point in $\sigma_j$ is denoted by $\tilde{\sigma}_{j,i}$, s.t.
$
  \sigma_j = \bigotimes_{i=1}^n \tilde{\sigma}_{j,i}
$.
When $A$ is a matrix, $e^A$ is the matrix exponential of $A$ defined
by
$
  e^A = \sum_{l=0}^\infty \frac{1}{l!} A^l
$
  .

\section{Simulated Annealing for Clustering} \label{sec:sa}

We briefly review simulated annealing (SA) \citep{Kirkpatrick83}
particularly for clustering.  SA is a stochastic optimization
algorithm.  An objective function is given as an energy function such
that a better solution has a lower energy.  In each step, SA searches
for the next random solution near the current one.  The next solution
is chosen with a probability that depends on temperature $T$ and on
the energy function value of the next solution.  SA almost randomly
choose the next solution when $T$ is high, and it goes down the hill
of the energy function when $T$ is low.  Slower cooling $T$ increases
the probability to find the global optimum.









Algorithm 1 summarizes a SA algorithm for clustering.
Given inverse temperature
$\beta = 1/T$, SA updates state $\sigma$ with,
\begin{align}
  p_\text{SA}(\sigma; \beta) = \frac{1}{Z} 
  \exp\left[ - \beta E(\sigma) \right]
  \label{eq:sa_model}
  ,
\end{align}
where $E(\sigma)$ is the energy function of state $\sigma$, and $Z$ is
a normalization factor defined by $Z =
\sum_{\sigma} \exp( - \beta E(\sigma) )$.  For probabilistic models,
the energy function is defined by $E(\sigma) \equiv - \log
p_\text{prob-model}(X,\sigma)$ where $p_\text{prob-model}(X,\sigma)$
is given by a probabilistic model and $X$ is data.  Note
$p_\text{SA}(\sigma;\beta=1)=p_\text{prob-model}(\sigma | X)$.  For
loss-function-based models (e.g. spectral clustering), which searches
for $\sigma = \argmin_\sigma loss(X,\sigma)$, the energy function is
defined by $E(\sigma) = loss(X,\sigma)$.

In many cases, the calculation of $Z$ in \eqref{eq:sa_model} is
intractable.  Thus, Markov chain Monte Carlo (MCMC) is utilized to
sample a new state from a current state.  In this paper, we focus on
the Gibbs sampler in MCMC methods.  Each step of the Gibbs sampler
draws an assignment of the $i$-th data point, $\tilde{\sigma}_i$, from,
\begin{align}
  &p_\text{SA}(\tilde{\sigma}_i | \sigma \backslash \tilde{\sigma}_i ) 
  =
  \frac{
    \exp\left[ -\beta E(\sigma) \right]
  }{
    \sum_{\tilde{\sigma}_i}
    \exp\left[ -\beta E(\sigma) \right]
  }
  ,
  \label{eq:sa_update}
\end{align}
where $\sigma \backslash \tilde{\sigma}_i$ means $\{  \tilde{\sigma}_j | j\neq i  \}$.
Note the denominator of \eqref{eq:sa_update} is summation over
$\tilde{\sigma}_i$, which is tractable ($\cO(k)$).

\section{Quantum Annealing for Clustering} \label{sec:qa}

Our goal of this section is to derive a sampling algorithm based on
quantum annealing (QA) for clustering.  On the way to the goal,
our contribution is three folds, which are a well-formed quantum
effect in Section \ref{sec:intro_qa}, an appropriate similarity
measure for clustering in Section \ref{sec:purity} and an annealing
schedule in Section \ref{sec:schedule}.

\subsection{Introducing Quantum Effect and the Suzuki-Trotter expansion}\label{sec:intro_qa}

Before we apply QA to clustering problems, we set
them up in a similar fashion to statistical mechanics.  We reformulate
\eqref{eq:sa_model} by the following equation,
\begin{align}
  p_\text{SA}(\sigma; \beta) = \frac{1}{Z} \sigma^T e^{-\beta\cH_\text{c}} \sigma
  \label{eq:sa_with_hamiltonian}
  ,
\end{align}
where $\cH_\text{c}$ is a $k^n$ by $k^n$ diagonal matrix.
$\cH_\text{c}$ is called (classical) Hamiltonian in physics.
For example, we have the following $\cH_\text{c}$ when $k=2$
and $n=2$.
\begin{align}
  \cH_\text{c} = \left(\begin{array}{cccc}
    \!\! E(\sigma^{(1)})\!\! & 0 & 0 & 0 \\
    0 & \!\!\!\!E(\sigma^{(2)})\!\! & 0 & 0 \\
    0 & 0 & \!\!\!\!E(\sigma^{(3)})\!\! & 0 \\
    0 & 0 & 0 & \!\!\!\!E(\sigma^{(4)})\!\!
\end{array}\right)
.
\end{align}
In this example, $\sigma^{(t)}$ indicates the $t$-th
assignment of $k^n$ available assignments, i,e.  $\sigma^{(1)} \!=\!
(1,~ 0,~ 0,~ 0)^T$, $\sigma^{(2)} \!=\! (0,~ 1,~ 0,~ 0)^T$,
$\sigma^{(3)} \!=\! (0,~ 0,~ 1,~ 0)^T$ and $\sigma^{(4)} \!=\! (0,~
0,~ 0,~ 1)^T$.  $e^{-\beta\cH_\text{c}}$ is the matrix exponential in
\eqref{eq:sa_with_hamiltonian}.  Since $\cH_\text{c}$ is diagonal,
$e^{-\beta\cH_\text{c}}$ is also diagonal with
$[e^{-\beta\cH_\text{c}}]_{tt} = \exp( -\beta E(\sigma^{(t)}) )$.
Hence, we find $\sigma^{(t)T} e^{-\beta\cH_\text{c}} \sigma^{(t)} =
\exp( -\beta E(\sigma^{(t)}))$ and \eqref{eq:sa_with_hamiltonian}
equal to \eqref{eq:sa_model}.
In practice, we use MCMC methods to sample $\sigma$ from
$p_\text{SA}(\sigma; \beta)$ in \eqref{eq:sa_with_hamiltonian} by
\eqref{eq:sa_update}.  This is because we do not need to calculate $Z$
and it is easy to evaluate $\sigma^{T} e^{-\beta\cH_\text{c}} \sigma$,
which is equal to $\exp( -\beta E(\sigma))$.


QA draws a sample from the following equation,
\begin{align}
  p_\text{QA}(\sigma; \beta, \Gamma) = \frac{1}{Z} \sigma^T e^{-\beta\cH} \sigma
  \label{eq:qa_model}
  ,
\end{align}
where $\cH$ is defined by
  $
  \cH = \cH_\text{c} + \cH_\text{q}
  $.
$\cH_\text{q}$ represents quantum effect.
We define $\cH_\text{q}$ by
  $\cH_\text{q} = \sum_{i=1}^n  \rho_i$ where
\begin{align}
  \rho_i \!=\! 
  \left(\bigotimes_{j=1}^{i-1} \bbE_k \!\right)
  \!\!\otimes
  \rho
  \otimes\!\!
  \left(\bigotimes_{j=i+1}^n \bbE_k \!\right)
  \!,~~~~
  \rho \!=\! \Gamma (\bbE_k - \bbone_k)
  \nn
  ,
\end{align}
$\bbE_k$ is the $k$ by $k$ identity matrix, and $\bbone_k$ is
the $k$ by $k$ matrix of ones, i.e. $[\bbone_k]_{ij} = 1$ for all $i$
and $j$.  For example, $\cH$ is,
\begin{align}
  \cH = \left(\begin{array}{cccc}
    \!\!\!\!E(\sigma^{(1)})\!\! & -\Gamma & -\Gamma & 0 \\
    -\Gamma & \!\!\!\!E(\sigma^{(2)})\!\! & 0 & -\Gamma \\
    -\Gamma & 0 & \!\!\!\!E(\sigma^{(3)})\!\! & -\Gamma \\
    0 & -\Gamma & -\Gamma & \!\!\!\!E(\sigma^{(4)})\!\!
\end{array}\right)
,
\end{align}
when $k=2$ and $n=2$.  The derived algorithm depends on quantum effect
$\cH_\text{q}$.  We found our definition of $\cH_\text{q}$ worked
well.  We also tried a couple of $\cH_\text{q}$.  We explain a bad
example of $\cH_\text{q}$ later in this section.

QA samples $\sigma$ from \eqref{eq:qa_model}.  For SA, MCMC methods
are exploited for sampling.  However, in quantum models, we cannot
apply MCMC methods directly to \eqref{eq:qa_model} because it is
intractable to evaluate $\sigma^T e^{-\beta\cH} \sigma$ unlike
$\sigma^{T} e^{-\beta\cH_\text{c}} \sigma = \exp( -\beta E(\sigma))$.
This is because $e^{-\beta\cH}$ is not diagonal whereas
$e^{-\beta\cH_\text{c}}$ is diagonal.  Thus, 
we exploit the Trotter product formula \citep{Trotter59} to
approximate \eqref{eq:qa_model}.
If $A_1,...,A_L$ are symmetric matrices, the Trotter product formula gives,
  $
  \exp\left( \sum_{l=1}^L A_l \right)
  \! = \!
  \left( \prod_{l=1}^L \exp(A_l/m) \right)^{\! m}
  \!\!\!\!\!
  + O\left(  \frac{1}{m}  \right)
  $.
Note the residual of finite $m$ is the order of $1/m$.  Hence, this
approximation becomes exact in the limit of $m\rightarrow \infty$.
Since $\cH = \cH_\text{c} + \cH_\text{q}$ is symmetric, we can apply the
Trotter product formula to \eqref{eq:qa_model}.
Following \citep{Suzuki76}, \eqref{eq:qa_model} reads the following expression,
\begin{theorem}
  \label{theorem:qa_is_approximated_by_st}
\begin{align}
  &p_\text{QA}(\sigma_1; \beta, \Gamma)
  \nn\\&\!\!\!=\!
  \sum_{\sigma_2}
  \!...\!\!
  \sum_{\sigma_m}
  p_\text{QA-ST}(\sigma_1,\sigma_2,...,\sigma_m; \beta, \Gamma)
  + O\!\left(  \frac{1}{m}  \right)
  \label{eq:qa_approximated_by_ST}
  ,
\end{align}
where
\begin{align}
  &p_\text{QA-ST}(\sigma_1,...,\sigma_m; \beta, \Gamma)
  \nn\\
  &
  \equiv
  \frac{1}{Z} 
  \prod_{j=1}^m
  p_\text{SA}(\sigma_j; \beta/m)
  e^{s(\sigma_j,\sigma_{j+1}) f(\beta,\Gamma)},
  \label{eq:qa_model_ST}
  \\
&s(\sigma_j,\sigma_{j+1}) =
\frac{1}{n}\sum_{i=1}^n \delta(\tilde{\sigma}_{j,i}, \tilde{\sigma}_{j+1,i}),
\label{eq:similarity}
\\ &
f(\beta, \Gamma) \!=\! n \log\!\!
\left(
  1+
\frac{ k }
{
e^{\frac{k\beta\Gamma}{m}} - 1
}
\right)
\label{eq:f}
\!.
\end{align}
\end{theorem}
The derivation from \eqref{eq:qa_model} to
\eqref{eq:qa_approximated_by_ST} is called the Suzuki-Trotter
expansion.  We show the details of the derivation in Appendix \ref{sec:appendix}.
\textit{\eqref{eq:qa_approximated_by_ST} means sampling $\sigma_1$
  from $p_\text{QA}(\sigma_1; \beta, \Gamma)$ is approximated by
  sampling $(\sigma_1,...,\sigma_m)$ from
  $p_\text{QA-ST}(\sigma_1,...,\sigma_m)$.}  \eqref{eq:qa_model_ST}
shows $p_\text{QA-ST}$ is similar to parallel $\{
p_\text{SA}(\sigma_j;\beta/m) \}_{j=1}^m$, but it has quantum
interaction $e^{s(\sigma_j,\sigma_{j+1})f(\beta,\Gamma)}$.  Note if
$f(\beta,\Gamma)=0$, i.e. $\Gamma=\infty$, the interaction disappears, and
$p_\text{QA-ST}$ becomes $m$ independent SAs.
$s(\sigma_j,\sigma_{j+1})$ takes $[0,1]$ where
$s(\sigma_j,\sigma_{j+1})=1$ when $\sigma_j=\sigma_{j+1}$ and
$s(\sigma_j,\sigma_{j+1})=0$ when $\sigma_j$ and $\sigma_{j+1}$ are
completely different.  Thus, we call $s(\sigma_j,\sigma_{j+1})$
similarity.
Even with finite $m$, we can show the approximation in
\eqref{eq:qa_approximated_by_ST} becomes exact after enough annealing
iterations has passed with our annealing schedule proposed in Section
\ref{sec:schedule}\footnote{The residual of the approximation in
  \eqref{eq:qa_approximated_by_ST} is dominated by $\beta^2\Gamma$
  with small $\beta$ and large $\Gamma$.  Using the annealing schedule
  proposed in Section \ref{sec:schedule}, the residual goes to zero as
  annealing continues ($\beta\rightarrow 0$ and $\Gamma\rightarrow\infty$).}.

The similarity in \eqref{eq:similarity} depends on quantum effect
$\cH_\text{q}$.  A different $\cH_\text{q}$ results in a different
similarity.  For example, we can derive an algorithm with quantum
effect $\cH_\text{q}' = \Gamma(\bbE_{k^n} - \bbone_{k^n})$.
$\cH_\text{q}'$ gives similarity $s'(\sigma_j,\sigma_{j+1}) =
\prod_{i=1}^n \delta(\tilde{\sigma}_{j,i}, \tilde{\sigma}_{j+1,i})$.
Going back to Fig.\ref{fig:local_global}, we notice
$s(\sigma_1,\sigma_2)>0$ but $s'(\sigma_1,\sigma_2)=0$.  In this case,
$p_\text{QA-ST}$ with $\cH_\text{q}'$ is just $m$ independent SAs
because interaction $f$ is canceled by $s'(\sigma_1,\sigma_2)=0$, and
$p_\text{QA-ST}$ is unlikely to search for $\sigma^*$.  On the other
hand, $p_\text{QA-ST}$ with $\cH_\text{q}$ is more likely to search
for $\sigma^*$ because interaction $f$ allows $\sigma_1$ and
$\sigma_2$ to go between $\sigma_1$ and $\sigma_2$.

Now, we can construct a Gibbs sampler based on $p_\text{QA-ST}$ in a
similar fashion to \eqref{eq:sa_update}.  Although the sampler is
tractable for statistical mechanics, it is intractable for machine
learning.  We give a solution to the problem in Section
\ref{sec:purity}.  We also discuss the annealing schedule of $\beta$
and $\Gamma$ in Section \ref{sec:schedule}, which is a crucial point
in practice.

\subsection{Cluster-Label Permutation} \label{sec:purity}

\begin{figure}[t!]
\begin{center}
  \begin{tabular}{cc}
    $\sigma_1=\sigma_2'$
    &
    $\sigma_2$
    \\
    \fbox{\includegraphics[width=.42\linewidth]{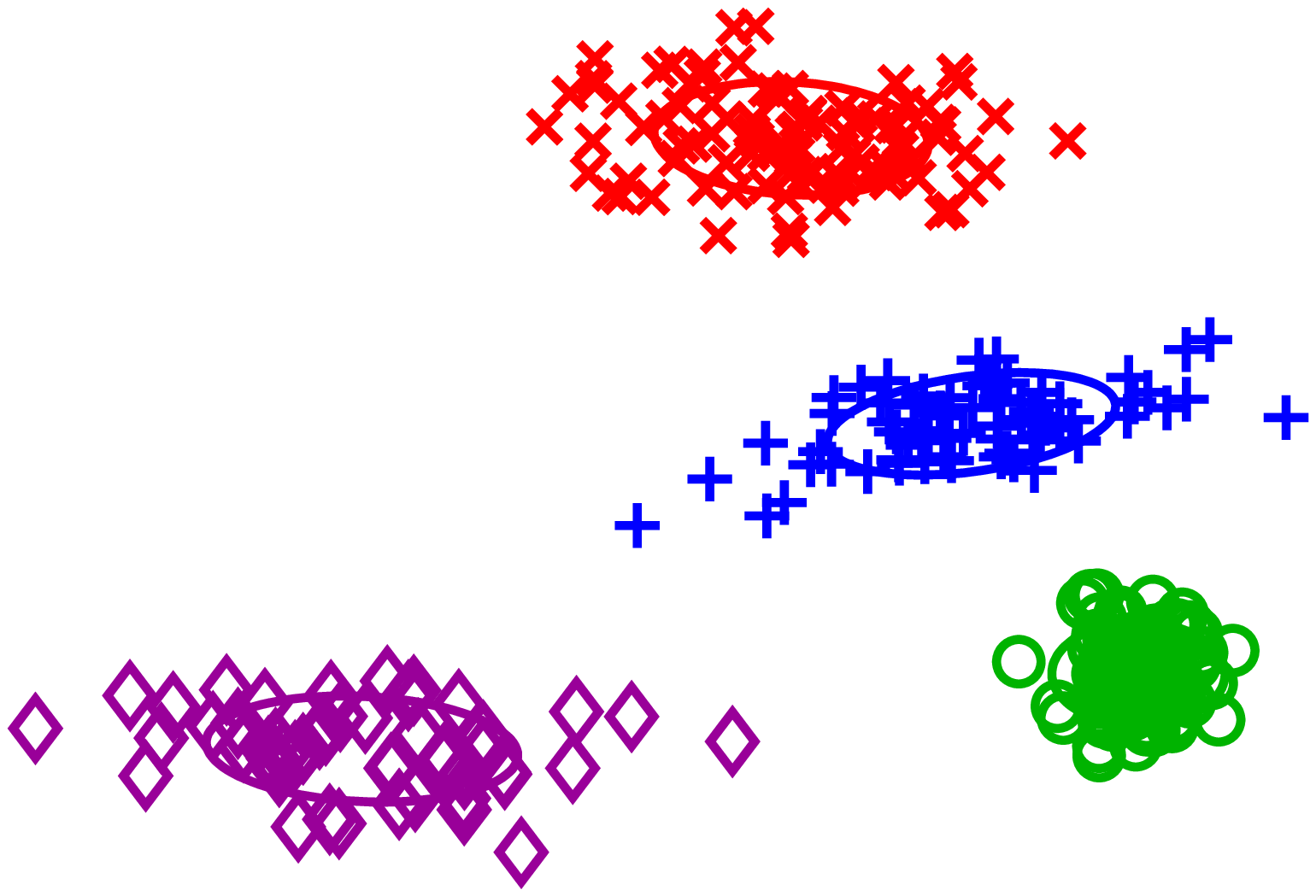}}
    &
    \fbox{\includegraphics[width=.42\linewidth]{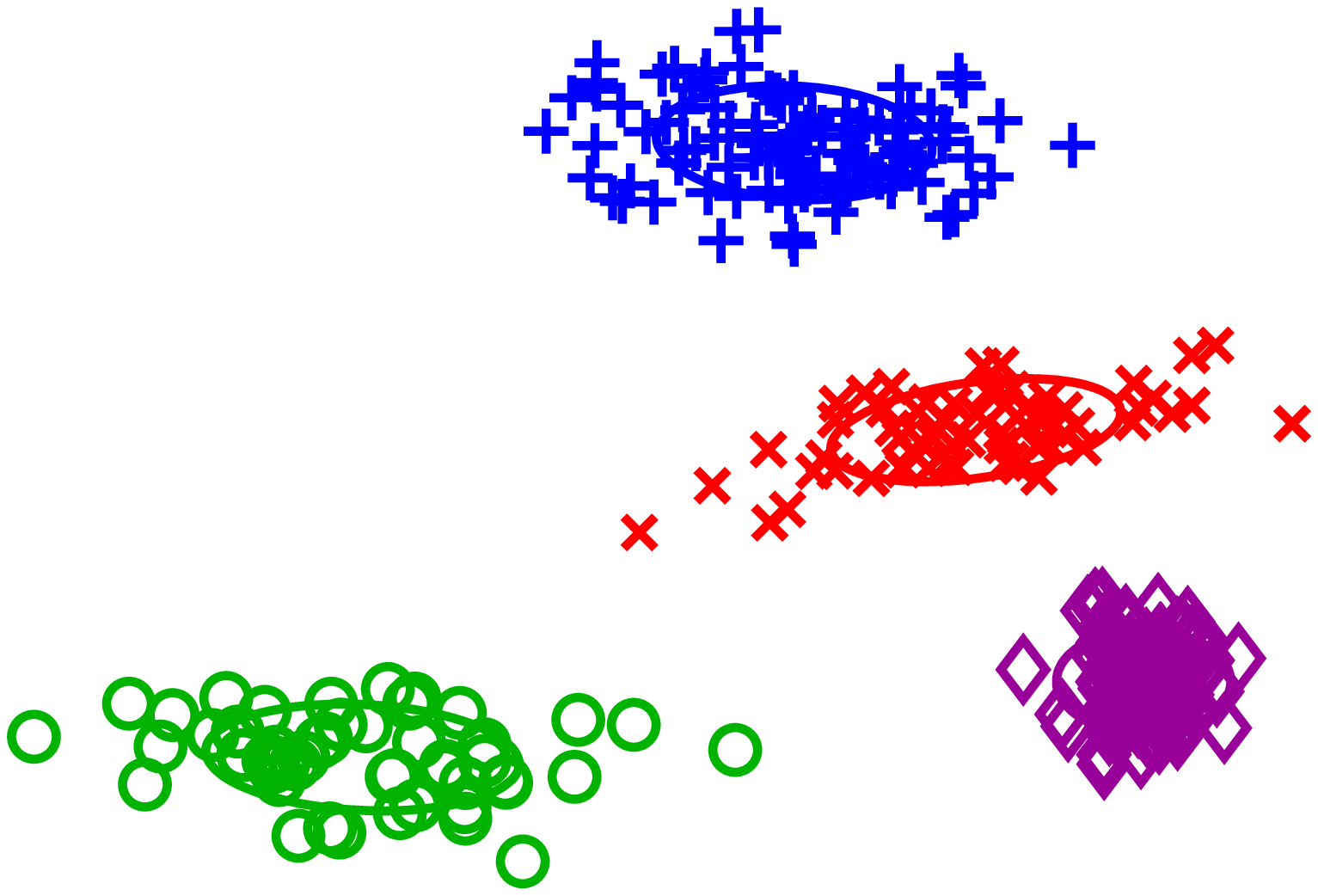}}
  \end{tabular}
\end{center}
\vspace{-1mm}
\caption{$\sigma_1$ and $\sigma_2$ give the same clustering but have
  different cluster labels. Thus, $s(\sigma_1,\sigma_2)=0$.  After
  cluster label permutation from $\sigma_2$ to $\sigma_2'$,
  $s(\sigma_1,\sigma_2')=1$.  The \textit{purity}, $\tilde{s}$, gives
  $\tilde{s}(\sigma_1,\sigma_2) = 1$ as well.}
\label{fig:purity}
\end{figure}
\begin{figure}[t!]
\begin{center}
\includegraphics[scale=0.6]{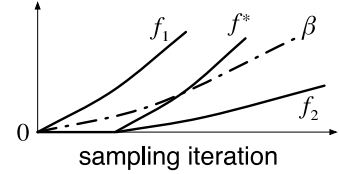}
\vspace{-4mm}
\end{center}
\vspace{-1mm}
\caption{The schedules of $\beta$ and $f(\beta,\Gamma)$.}
\label{fig:beta_f}
\vspace{-3mm}
\end{figure}

Our goal is to make an efficient sampling algorithm.  In a similar
fashion to \eqref{eq:sa_update}, we can construct a Gibbs sampler
$p_\text{QA-ST}(\tilde{\sigma}_{j,i} | \{\sigma\}_{j=1}^m \backslash
\tilde{\sigma}_{j,i})$ whose computational complexity is the same as
that of \eqref{eq:sa_update}.
However, the sampler can easily get stuck in local optima, which is for example 
$p_\text{QA-ST}(\sigma_1,\sigma_2)$ in Fig.\ref{fig:purity}.  If we
can draw $\sigma_2'$ in Fig.\ref{fig:purity} from $\sigma_2$,
$p_\text{QA-ST}(\sigma_1,\sigma_2')$ is a better state than
$p_\text{QA-ST}(\sigma_1,\sigma_2)$ i.e.
$p_\text{QA-ST}(\sigma_1,\sigma_2') \geq
p_\text{QA-ST}(\sigma_1,\sigma_2)$ because $s(\sigma_1,\sigma_2') =
1$, $s(\sigma_1,\sigma_2) = 0$ and $f(\beta,\Gamma) \geq 0$ in
\eqref{eq:qa_model_ST}.  Since sampler $p_\text{QA-ST}(\tilde{\sigma}_{j,i} |
\{\sigma\}_{j=1}^m \backslash \tilde{\sigma}_{j,i})$ only changes the
label of one data point at a time, the sampler cannot sample
$\sigma_2'$ from $\sigma_2$ efficiently.  In statistical mechanics, a
cluster label permutation sampler is applied to cases such as
Fig.\ref{fig:purity}.  The label permutation sampler does not change
cluster assignments but draws cluster label permutation, e.g. $\sigma_2'$ from
$\sigma_2$ in one step.
In other words, the sampler exchanges rows of matrix $Y(\sigma)$ defined in \eqref{eq:Y}.
In the case of Fig.\ref{fig:purity}, $k$ is
equal to four, so we have $4!=24$ choices of label permutation.  The
computational complexity of the sampler is $\cO(k!)$ because its
normalization factor requires summation over $k!$
choices.  The sampler is tractable for statistical mechanics due to
relatively small $k$.  However, it is intractable for machine learning
where $k$ can be very large.

We introduce approximation of $p_\text{QA-ST}$ so that we do not need
to sample label permutation, whose computational complexity is
$\cO(k!)$.  In particular, we replace similarity
$s(\sigma_j,\sigma_{j+1})$ in \eqref{eq:qa_model_ST} by the
\textit{purity}, $\tilde{s}(\sigma_j,\sigma_{j+1})$.  The
\textit{purity}, $\tilde{s}(\sigma_j,\sigma_{j+1})$, is defined by
$
\tilde{s}(\sigma_j,\sigma_{j+1})
\equiv \frac{1}{n} \sum_{c=1}^k \max_{c'=1...k} \left[  Y(\sigma_j)Y(\sigma_{j+1})^T  \right]_{c,c'}
$
where $Y$ is defined in \eqref{eq:Y}, and $[A]_{c,c'}$ denotes the
$(c,c')$ element of matrix $A$.
In the case of Fig.\ref{fig:purity}, 
$\tilde{s}(\sigma_1,\sigma_{2}) = 1$ whereas
$s(\sigma_1, \sigma_{2}) = 0$.  In general,
$s(\sigma_1, \sigma_{2}) \leq \tilde{s}(\sigma_1, \sigma_{2})$.

Let $\tilde{\sigma}_{j,i}$ be the $i$-th data point of assignment $\sigma_j$.
The update probability of
$\tilde{\sigma}_{j,i}$ with the \textit{purity}
is,
\begin{align}
   &p_\text{QA-ST+\textit{purity}}(\tilde{\sigma}_{j,i} | \{\sigma_j\}_{j=1}^m \backslash \tilde{\sigma}_{j,i}; \beta, \Gamma)
  \nn\\ 
  & \!\!
  =
  \frac{
    \exp\left[
      -\frac{\beta}{m} E(\sigma_j)
      +
      \tilde{s}(\sigma_{j-1},\sigma_j,\sigma_{j+1})
      f(\beta, \Gamma)
    \right]
  }{
    \underset{{\tilde{\sigma}_{j,i}}}{\sum}
    \exp\!\!\left[\!
      -\frac{\beta}{m} E(\sigma_j)
      +
      \tilde{s}(\sigma_{j-1},\sigma_j,\sigma_{j+1})
      f(\beta, \Gamma)
    \!\right]
  }
  \label{eq:qa_update}
  ,
\end{align}
where $ \tilde{s}(\sigma_{j-1},\sigma_j,\sigma_{j+1}) =
\tilde{s}(\sigma_{j},\sigma_{j-1}) +
\tilde{s}(\sigma_{j},\sigma_{j+1}) $\footnote{Note $\tilde{s}$ is not
  commutative, and take care of the order of the arguments of
  $\tilde{s}$.  We use $\tilde{s}(\sigma_{j-1},\sigma_j,\sigma_{j+1})
  = \tilde{s}(\sigma_{j},\sigma_{j-1}) +
  \tilde{s}(\sigma_{j},\sigma_{j+1})$, but we omit the reason due to
  space.  }.
The computational complexity of
\eqref{eq:qa_update} is $\cO(k^2)$, but caching statistics reduces it
to $\cO(k)$.  Thus, Step \ref{alg:qa:init} in Algorithm 2 requires
$\cO(k^2)$, and Step \ref{alg:qa:draw} requires $\cO(k)$, which is the
same as SA.

Using another representation of $\sigma_j$ and a different
$\cH_\text{q}$, we can develop a sampler, which does not need label
permutation.  However, its computational complexity is $\cO(n)$ in
Step \ref{alg:qa:draw}, which is much more expensive than $\cO(k)$.
Thus, the sampler is less efficient than the proposed sampler even
though the sampler does not need to solve label permutation.

\subsection{Annealing Schedule of $\beta$ and $\Gamma$} \label{sec:schedule}

The annealing schedules of $\beta$ and $\Gamma$ significantly affect
the result of QA.  Thus, it is crucial to use good schedules of
$\beta$ and $\Gamma$.  In this section, we propose the
annealing schedule of $\Gamma$ and $\beta$.

\begin{algorithm}[t!]
  \caption{Simulated Annealing for Clustering}
  \begin{algorithmic}[1]
    \State Initialize inverse temperature $\beta$ and assignment $\sigma$.

    \Repeat

    \For{$i=1,...,n$}

    \State Draw the new assignment of the $i$-th data point,
    $\tilde{\sigma}_i$, with a probability given in
    \eqref{eq:sa_update}.

    \EndFor

    \State Increase inverse temperature $\beta$.

    \Until{State $\sigma$ converges} 
  \end{algorithmic}
  \label{alg:sa}
\end{algorithm}

\begin{algorithm}[t!]
  \caption{Quantum Annealing for Clustering}
  \begin{algorithmic}[1]
    \State Initialize inverse temperature $\beta$ and quantum annealing parameter
    $\Gamma$. \label{alg:qa:init}



    \Repeat

    \For{$j = 1,...,m$}

    \For{$i=1,...,n$}

    \State Draw the new assignment of the $i$-th data point,
    $\sigma_{j,i}$, with a probability given in \eqref{eq:qa_update}.
    \label{alg:qa:draw}

    \EndFor

    \EndFor

    \State Increase inverse temperature $\beta$, and decrease QA
    parameter $\Gamma$.\label{alg:qa:update_beta_Gamma}

    \Until{State $\sigma$ converges} 
  \end{algorithmic}
  \label{alg:qa}
\end{algorithm}

We address two points before proposing a schedule.  One is our
observation from pilot experiments, and the other is the balance of
$\beta$ and $\Gamma$.  From our pilot experiments, we observe QA-ST works
well when it can find suboptimal assignments $\{\sigma_j\}_{j=1}^m$ by
convergence.  \eqref{eq:qa_update} shows QA-ST searches for a better
assignment from suboptimal $\{\sigma_j\}_{j=1}^m$.  On the other hand,
when current $\{\sigma_j\}_{j=1}^m$ are far away from global optimum
or even sub-optima, QA-ST does not necessarily work well.  
Comparing \eqref{eq:qa_update} with \eqref{eq:sa_update} in terms of
$\beta$ and $\Gamma$, if $\frac{\beta}{m} \gg f(\beta,\Gamma)$,
$\{\sigma_j\}_{j=1}^m$ are sampled from $p_\text{SA}(\sigma_j)$,
i.e. no interaction between $\sigma_j$ and $\sigma_{j+1}$.  On the
other hand, if $\frac{\beta}{m} \ll f(\beta,\Gamma)$, $\{\sigma_j\}_{j=1}^m$
become very close to each other regardless of energy $E(\sigma_j)$.

From the above discussion,
$\beta/m$
at the beginning should be larger
than $f(\beta,\Gamma)$ and large enough to collect suboptimal
assignments, and $f(\beta,\Gamma)$ should become larger than
$\beta/m$
at some point to make $\{\sigma_j\}_{j=1}^m$ closer.  The best path of
$\beta$ and $f(\beta,\Gamma)$ would be like $f^*$ in
Fig.\ref{fig:beta_f}.  $f_1$ in Fig.\ref{fig:beta_f} is stronger than
$\beta$ from the beginning, which does not allow QA-ST to search for good
assignments due to too strong quantum interaction
$f(\beta,\Gamma)\tilde{s}(\sigma_{j-1},\sigma_j,\sigma_{j+1})$ in
\eqref{eq:qa_update}.  $f_2$ is always smaller than $\beta$.  Hence,
QA-ST never makes $\{\sigma_j\}_{j=1}^m$ closer.  In other words, QA-ST does
not search for a middle (hopefully better) assignment from
$\{\sigma_j\}_{j=1}^m$.
Specifically, we use the following annealing schedule in Step
\ref{alg:qa:update_beta_Gamma} in Algorithm 2.
\begin{align}
  \beta = \beta_0 r_\beta^i,
  ~~~~~~
  \Gamma = \Gamma_0 \exp( - r_\Gamma^i )
  \label{eq:schedule_gamma}
  ,
\end{align}
where $r_\beta$ and $r_\Gamma$ are constants and $i$ denotes the
$i$-th iteration of sampling.  \eqref{eq:schedule_gamma} comes from the
following analysis of $f(\beta,\Gamma)$.  When $\frac{k\beta\Gamma}{m} \ll 1$,
\eqref{eq:f} reads,
  $
f(\beta, \Gamma)
\approx
-
n \log\left(
  \frac{\beta\Gamma}{m}
  \right)
=
n r_\Gamma^i
- 
n \log\left(
  \frac{\beta\Gamma_0}{m}
  \right)
  \nn
  $
.
Thus, the path of $f(\beta,\Gamma)$ become $f^*$ in
Fig.\ref{fig:beta_f} when $\Gamma_0$ is large enough and $r_\beta <
r_\Gamma$.  In this paper\footnote{
  When QA-ST is applied
  to loss-function-based models, ``until $\beta=m$'' should be
  calibrated according to loss-functions.},
 we set $\Gamma_0$ to a large value such
that $f(\beta,\Gamma)\approx 0$ until $\beta=m$.  This means
$p_\text{QA-ST}(\sigma_1,...,\sigma_m)$ is just $m$ independent instances of 
$p_\text{SA}(\sigma_j; \beta/m)$ until $\beta=m$.

Note there is not much difference of difficulty between SA and QA-ST
to choose annealing schedules.  In general, we should choose the
schedule of $\Gamma$ to be $f^*$ when the schedule of $\beta$ is
given.  As shown in the next section, QA-ST works well with $r_\Gamma
\approx r_\beta \times 1.05$.  Thus, the difficulty of choosing
annealing schedules for QA-ST is reduced to that of choosing the
schedule of $\beta$ for SA.


\begin{figure*}[t!]
\begin{center}

\includegraphics[scale=.39]{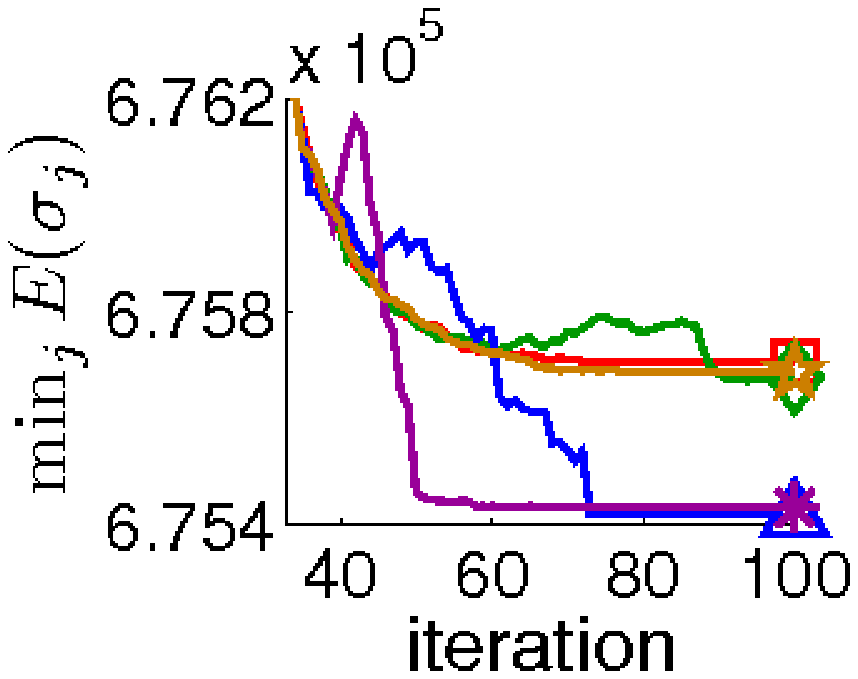}
\hspace{3mm}
\includegraphics[scale=0.39]{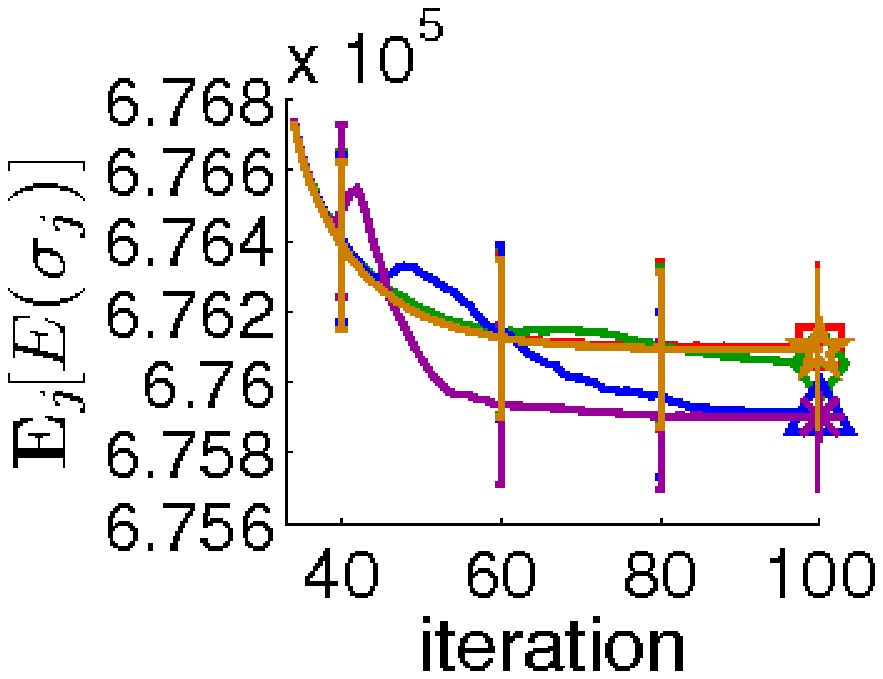}
\hspace{3mm}
\includegraphics[scale=.39]{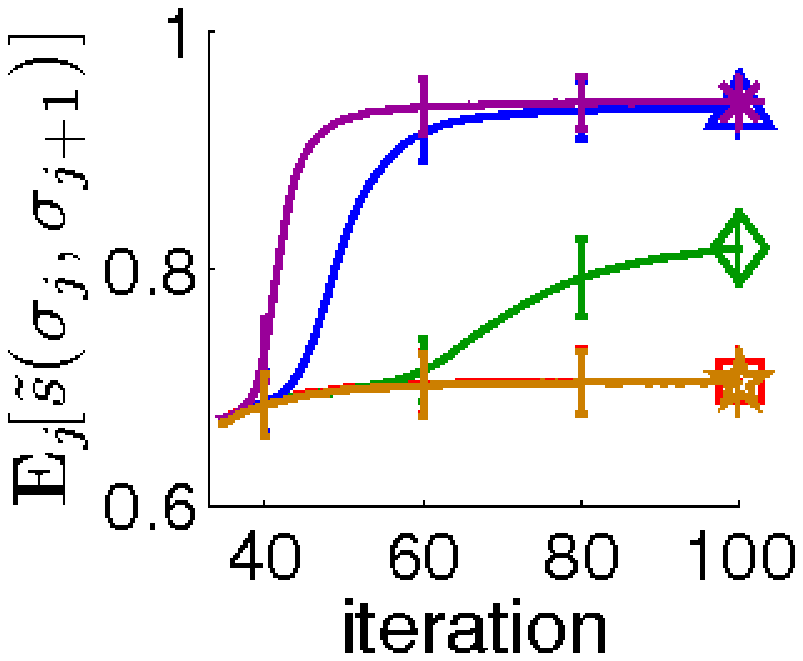}
\hspace{-1mm}
\begin{minipage}[b]{.18\linewidth}
  ~{\small MNIST with MoG}
  \\
  ~$r_\beta = 1.05$
  \vspace{-1mm}
  \\
\includegraphics[scale=.39]{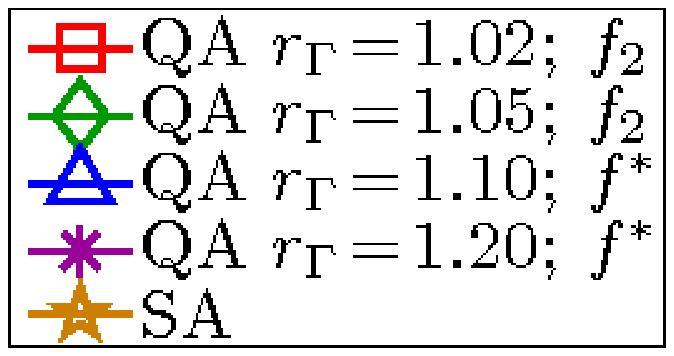}
\end{minipage}

\includegraphics[scale=.39]{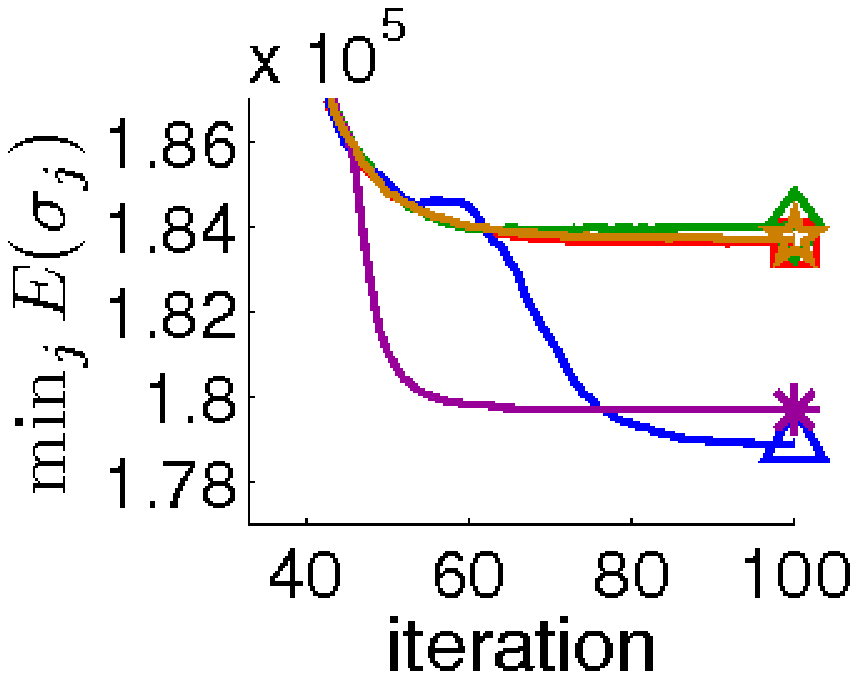}
\hspace{3mm}
\includegraphics[scale=.39]{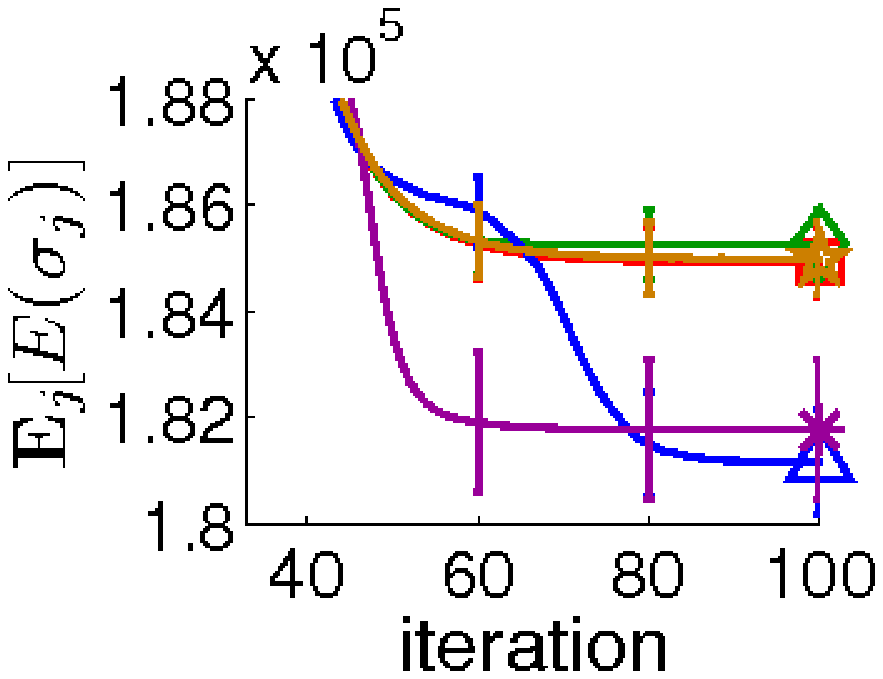}
\hspace{3mm}
\includegraphics[scale=.39]{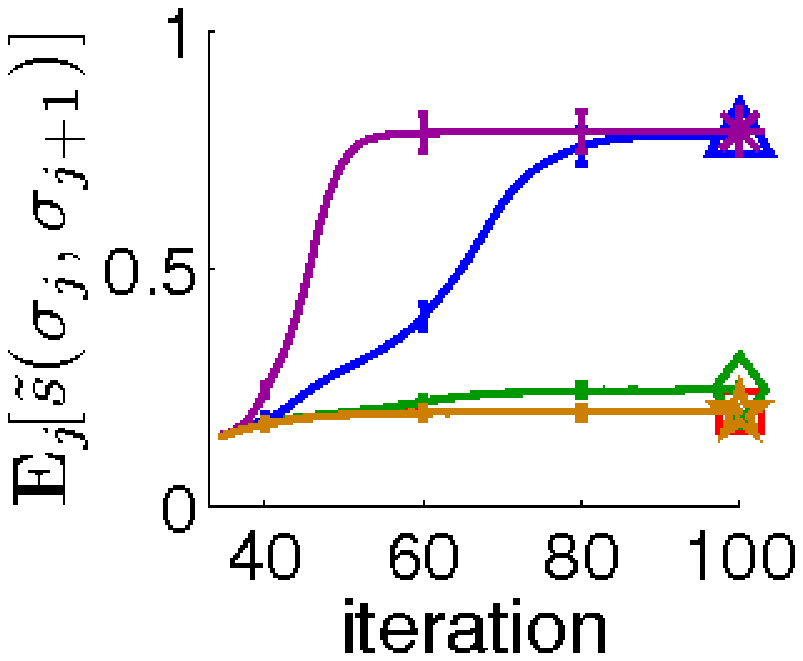}
\hspace{-1mm}
\begin{minipage}[b]{.18\linewidth}
  ~{\small Reuters with LDA}
  \\
  ~$r_\beta = 1.05$
  \vspace{-1mm}
  \\
\includegraphics[scale=.39]{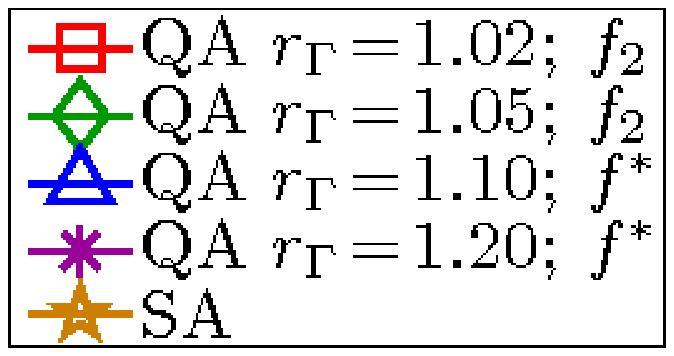}
\end{minipage}

\includegraphics[scale=.39]{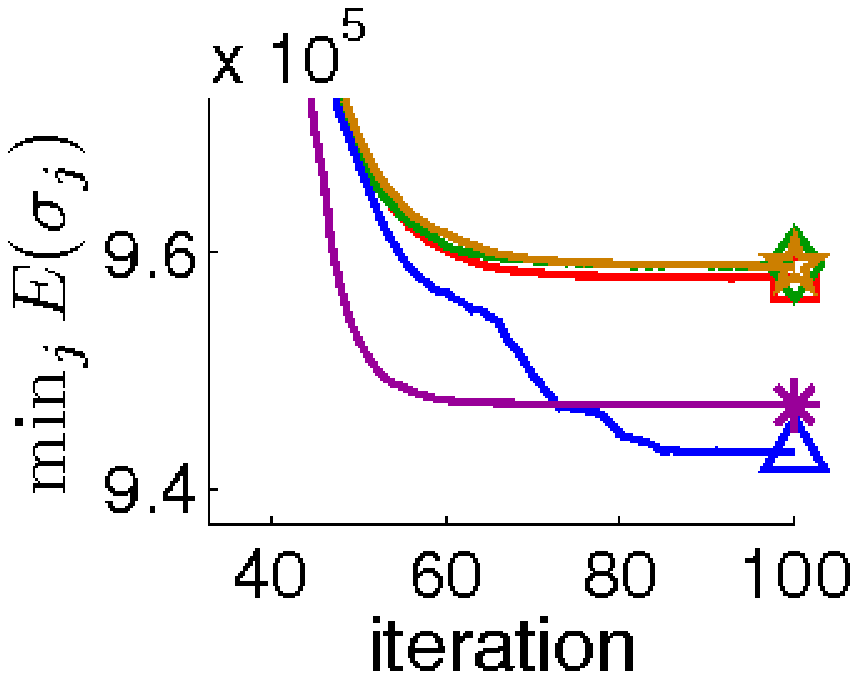}
\hspace{3mm}
\includegraphics[scale=.39]{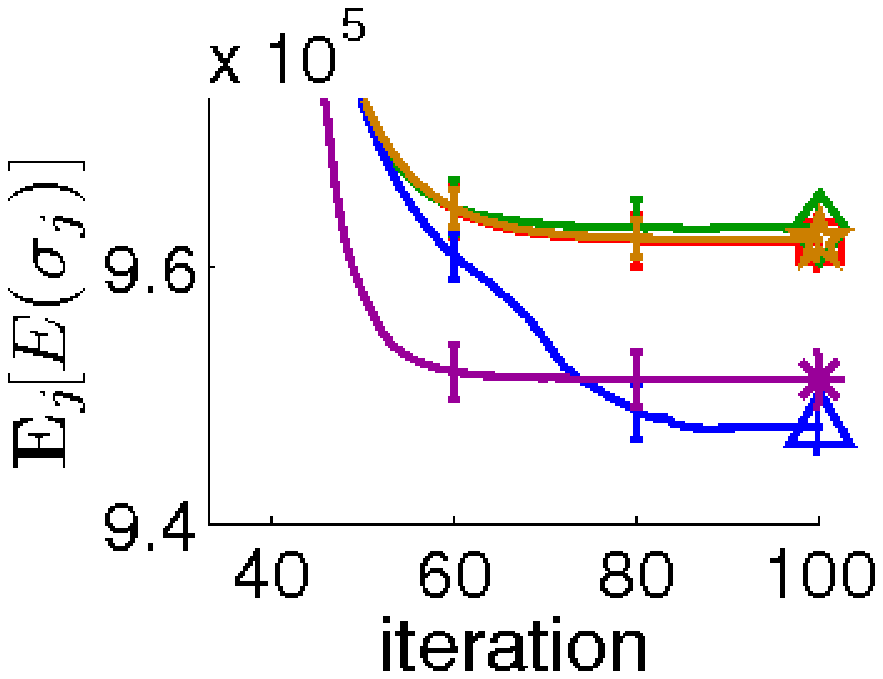}
\hspace{3mm}
\includegraphics[scale=.39]{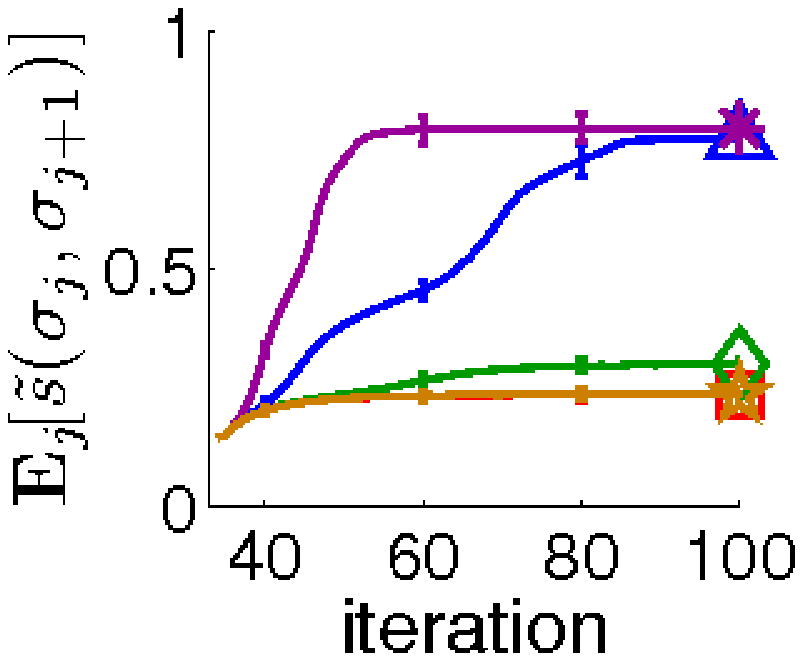}
\hspace{-1mm}
\begin{minipage}[b]{.18\linewidth}
  ~NIPS with LDA
  \\
  ~$r_\beta = 1.05$
  \vspace{-1mm}
  \\
\includegraphics[scale=.39]{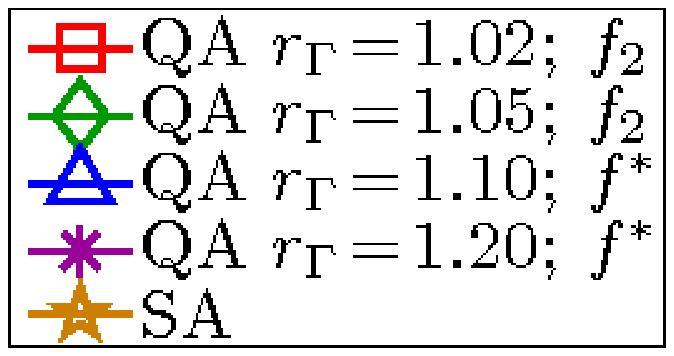}
\end{minipage}

\includegraphics[scale=.39]{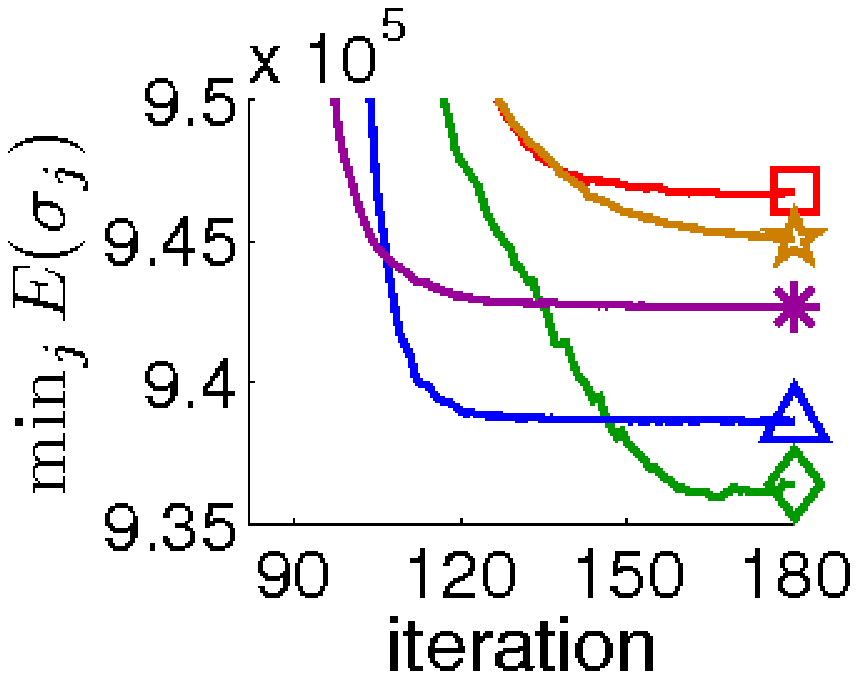}
\hspace{3mm}
\includegraphics[scale=.39]{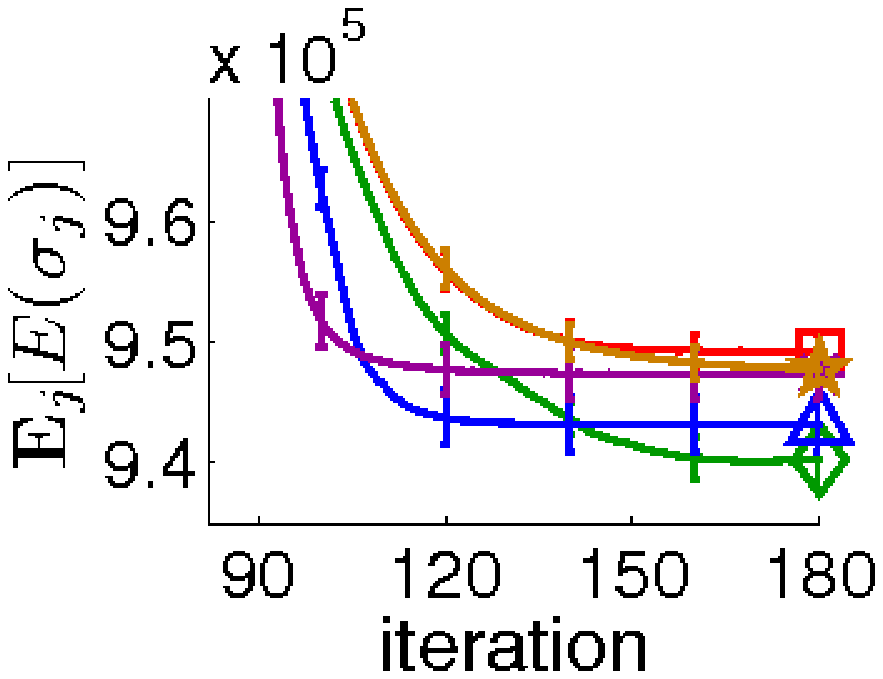}
\hspace{3mm}
\includegraphics[scale=.39]{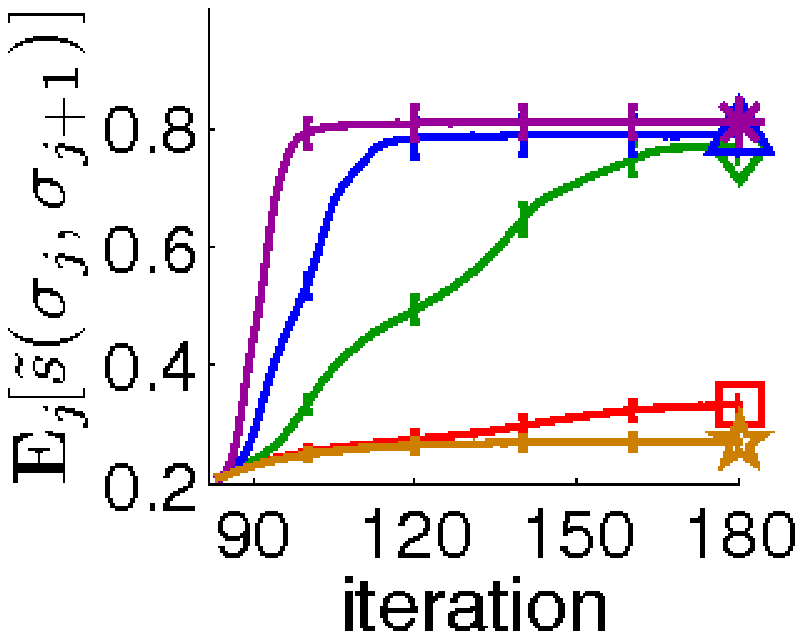}
\hspace{-1mm}
\begin{minipage}[b]{.18\linewidth}
  ~NIPS with LDA
  \\
  ~$r_\beta = 1.02$
  \vspace{-1mm}
  \\
\includegraphics[scale=.39]{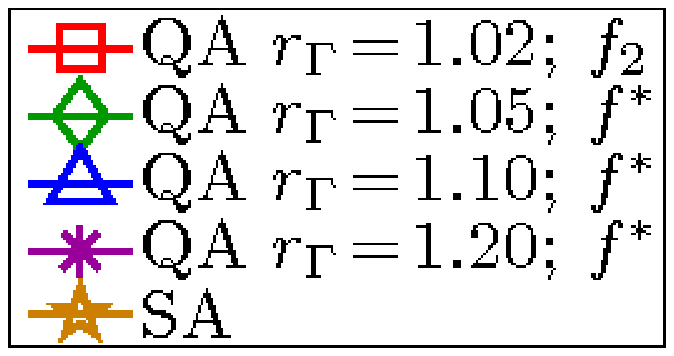}
\end{minipage}

\vspace{-3mm}
\end{center}
\caption{Comparison between SA and QA varying annealing schedule.  $r_\Gamma$,
  $f_2$ and $f^*$ in legends correspond to Fig.\ref{fig:beta_f}.
  The left-most column shows what SA and QA found.  QA with $f^*$ always found
  better results than SA.
}
\label{fig:experiments}
\end{figure*}

\section{Experiments} \label{sec:experiments}

We show experimental results in Fig.\ref{fig:experiments}.
In the top three rows of Fig.\ref{fig:experiments}, we vary the schedule of $\Gamma$ with a
fixed schedule of $\beta$ to see QA-ST work better than the best energy of $m$ SAs when the
schedule of $\Gamma$ lets the path of $f(\beta,\Gamma)$ be $f^*$ in
Fig.\ref{fig:beta_f}.  In the bottom row of the figure, we
compare QA-ST and SA with a slower schedule of $\beta$.
This experiment shows whether QA-ST still
works better than SA or not while the slower schedule of $\beta$
improves SA.

We apply SA and QA-ST to two models, which are a mixture of Gaussians
(MoG) with a conjugate normal-inverse-Wishart prior and the latent
Dirichlet allocation (LDA) \citep{Blei03}.  For both models, parameters
are marginalized out, and $E(\sigma) \equiv - \log p(X,\sigma)$ where
$X$ is data.  Thus, QA-ST and SA search for maximum a posteriori (MAP)
assignment $\sigma$.  MoG is applied to MNIST data set,
and LDA is applied to NIPS
corpus and
Reuters.
For MNIST, we
randomly choose 5,000 data points and apply PCA to reduce the
dimensionality to 20.  NIPS corpus has 1,684 documents, and we randomly choose 1000
words in vocabulary.  We also randomly choose 1000 documents and 2000 words in vocabulary 
from Reuters.  We set
$k$ to 30, 20 and 20 for MNIST, NIPS corpus and Reuters,
respectively.  We use the same schedule of $\beta$ for QA-ST and SA.  In
particular, we use the same $r_\beta$ for QA-ST and SA, and we set
$\beta_0=0.2$ for SA and $\beta_0=0.2m$ for QA-ST.  The difference of
$\beta_0$ for SA and QA-ST keeps QA-ST similar to SA in terms of
$\beta$-annealing for fair comparison (see \eqref{eq:sa_update} and
\eqref{eq:qa_update}).  For each data, we vary $r_\Gamma$ from 1.02 to 1.20 with fixed
$r_\beta$.  When $r_\beta \leq r_\Gamma$, the path of
$f(\beta,\Gamma)$ becomes $f^*$ in Fig.\ref{fig:beta_f}.
For any data set, $m$ is set to 50 for QA-ST.
We set $m$ of SAs so that $m$ SAs consume the same time as QA-ST.
Thus, we can compare QA-ST and $m$ SAs in terms of iteration in
Fig.\ref{fig:experiments}.
Consequently, $m$ of SA was set to 51, 55 and 55 for MNIST, NIPS and
Reuters, respectively\footnote{QA-ST and $m$ SAs took 21.7 and 22.0
  hours for MNIST, 62.5 and 62.8 hours for NIPS and 9.9 and 10.0 hours
  for Reuters.}.  In Fig.\ref{fig:experiments},
we plot only after
$\beta=m$ for QA-ST and $\beta=1$ for SA, which happen at the same
iteration for QA-ST and SA.

In Fig.\ref{fig:experiments}, the left column and the middle column
show the minimum and the mean energy of $\{ \sigma_j \}_{j=1}^m$.
Since this is an optimization problem, we are interested in the
minimum energy in the left column.  For each data, QA-ST with $f^*$
achieved better results than SA.  The right column of
Fig.\ref{fig:experiments} shows the mean of purity $\tilde{s}$.  As we
expect, the larger $r_\Gamma$ resulted in the higher $\tilde{s}$.
The bottom row of Fig.\ref{fig:experiments} shows the result of NIPS with
the slower schedule of $\beta$ than the schedule in the third row of 
Fig.\ref{fig:experiments}.  Although SA found better results 
than the third row of Fig.\ref{fig:experiments}, QA-ST
still worked better than SA.
Our experimental results are also consistent with the claim of
\citet{matsuda08}, which is that QA works the better than SA for more
difficult problems.  QA worked better for LDA than MoG.  The right
column of Fig.\ref{fig:experiments} shows $\tilde{s}$ of LDA converged
to smaller values than that of MoG.  This means LDA has more local
optima than MoG.

In this section, we have shown QA-ST achieves better results than SA when
the schedule of $\Gamma$ is $f^*$ in Fig.\ref{fig:beta_f}.  We
have also shown even with the slower schedule of $\beta$, QA-ST still
works better than SA.


\section{Discussion \& Conclusion} \label{sec:conclusion}



Many techniques to accelerate sampling have been studied.  Such
techniques can be applied to the proposed algorithm.  For example, the
split-merge sampler \citep{Richardson97} and the permutation augmented
sampler \citep{Liang07} use a global move to escape from local minima.
These techniques are available for the proposed algorithm as well.  We
can also apply the exchange Monte Carlo method.

We have applied quantum annealing (QA) to clustering.  To our best
knowledge, this is the first study of QA for clustering.  We have
proposed quantum effect $\cH_\text{q}$ fit to clustering and derived a
QA-based sampling algorithm.  We have also proposed a good annealing
schedule for QA, which is crucial for applications.  The computational
complexity of QA is larger than a single simulated annealing (SA).
However, we have empirically shown QA finds a better clustering
assignment than the best one of multiple-run SAs that are randomly
restarted until they consumes the same time as QA.  In other words, QA
is better than SA when we run SA many times. Actually, it is typical
to run SA many times because SA's fast cooling schedule of temperature
$T$ does not necessarily find the global optimum.  Thus, we strongly
believe QA is a novel alternative to SA for optimizing clustering.  In
addition, it is easy to implement the proposed algorithm because it is
very similar to SA.


Unfortunately, there is no proof yet that QA is better than SA in
general.  Thus, we need to experimentally show QA's performance for
each problem like this paper.  However, it is worth trying to develop
QA-based algorithms for different models, e.g. Bayesian networks, by
different quantum effect $\cH_\text{q}$.
The proposed algorithm looks like genetic algorithms in terms of
running multiple instances.  Studying their relationship is also
interesting future work.

{\small
\subsubsection*{Acknowledgements} 
  \vspace{-2mm}
This work was partially supported by
Research on Priority Areas
``Physics of new quantum phases in superclean materials'' (Grant No.
17071011) from MEXT, and also by the Next Generation Super Computer
Project, Nanoscience Program from MEXT.
Special thanks to Taiji Suzuki, T-PRIMAL members and Sato Lab.
}

\appendix
\section{The Details of the Suzuki-Trotter Expansion}\label{sec:appendix}
Following \citet{Suzuki76}, we give the details of derivation of
Theorem \ref{theorem:qa_is_approximated_by_st}.
The following Trotter product formula \citep{Trotter59}
says
if $A_1,\cdots,A_n$ are symmetric matrices, we have
\vspace{-1mm}
\begin{align}
  \exp\left( \sum_{l=1}^L A_l \right)
  \! = \!
  \left( \prod_{l=1}^L \exp(A_l/m) \right)^{\! m}
  \!\!\!\!\!
  + O\left(  \frac{1}{m}  \right)
  .
\end{align}
Applying the Trotter product formula to \eqref{eq:qa_model},
we have
\vspace{-1mm}
\begin{align}
  &
  p_\text{QA}(\sigma_1; \beta, \Gamma)
  =
  \frac{1}{Z}
  \sigma_1^T e^{- \beta (\cH_\text{c} + \cH_\text{q})} \sigma_1
  \nn\\
  &
  ~~~~~
  =
  \frac{1}{Z}
  \sigma_1^T \left(
  e^{- \frac{\beta}{m} \cH_\text{c}} 
  e^{- \frac{\beta}{m} \cH_\text{q}} 
  \right)^m
  \sigma_1
  + O\left(  \frac{1}{m}  \right)
  \label{eq:trotter_expansion}
  .
\end{align}
Note
\vspace{-1mm}
\begin{align}
\sigma_1^T \!\left(e^A\right)^{\!2}\! \sigma_1
&= \sigma_1^T e^A \bbE_k  e^A \sigma_1
= \sigma_1^T e^A \left( \sum_{\sigma_2} \sigma_2 \sigma_2^T \right) e^A \sigma_1
\nn\\
&= \sum_{\sigma_2} \sigma_1^T e^A \sigma_2 \sigma_2^T e^A \sigma_1
.
\end{align}
Hence, we express
\eqref{eq:trotter_expansion} by marginalizing out auxiliary variables
$\{\sigma_1',\sigma_2,\sigma_2',...,\sigma_m,\sigma_m'\}$,
\begin{align}
  &
  \frac{1}{Z}
  \sigma_1^T
  \left(
  e^{- \frac{\beta}{m} \cH_\text{c}}
  e^{- \frac{\beta}{m} \cH_\text{q}} 
  \right)^m
  \sigma_1
  \label{eq:trotter_expansion_inner}
  \\
  &=
  \frac{1}{Z}
  \sum_{\sigma_1'}
  \sum_{\sigma_2}
  ...
  \sum_{\sigma_m}
  \sum_{\sigma_m'}
  \sigma_1^T
  e^{- \frac{\beta}{m} \cH_\text{c}}
  \sigma_1'
  \sigma_1'^T
  e^{- \frac{\beta}{m} \cH_\text{q}} 
  \sigma_{2}
  \times ...
  \nn \\ &~~~~~~~~~~~~
  \times
  \sigma_m^T
  e^{- \frac{\beta}{m} \cH_\text{c}}
  \sigma_m'
  \sigma_m'^T
  e^{- \frac{\beta}{m} \cH_\text{q}} 
  \sigma_{m+1}
  \\
  &=
  \frac{1}{Z}
  \sum_{\sigma_1'}
  \sum_{\sigma_2}
  ...
  \sum_{\sigma_m}
  \sum_{\sigma_m'}
  \prod_{j=1}^m
  \sigma_j^T
  e^{- \frac{\beta}{m} \cH_\text{c}}
  \sigma_j'
  \sigma_j'^T
  e^{- \frac{\beta}{m} \cH_\text{q}}
  \sigma_{j+1}
  ,
  \label{eq:qa_naive_st_expansion}
\end{align}
where $\sigma_{m+1} = \sigma_1$.
To simplify \eqref{eq:qa_naive_st_expansion} more, we use the
following Lemma \ref{lemma:trace_classical} and Lemma \ref{lemma:trace_quantum}.

\begin{lemma}
  \label{lemma:trace_classical}
\begin{align}
  \sigma_j^T
  e^{- \frac{\beta}{m} \cH_\text{c}}
  \sigma_j'
  =&
  \exp\left(- \frac{\beta}{m} E(\sigma_j) \right)
  \delta(\sigma_j, \sigma_j')
  \nn\\
  \propto&
  p_\text{SA}(\sigma_j; \beta/m)
  \delta(\sigma_j, \sigma_j')
  \label{eq:trace_classical}
  ,
\end{align}
where $\delta(\sigma_j, \sigma_j') = 1$ if $\sigma_j=\sigma_j'$ and
$\delta(\sigma_j, \sigma_j') = 0$ otherwise.
\end{lemma}
\begin{proof}
  By the definition, $e^{- \frac{\beta}{m} \cH_\text{c}}$ is diagonal
  with $[e^{- \frac{\beta}{m} \cH_\text{c}}]_{tt} = E(\sigma^{(t)})$,
  and $\sigma_j$ and $\sigma_j'$ are binary indicator vectors,
  i.e. only one element in $\sigma_j$ is one and the others are zero.
  Thus, the above lemma holds.
\end{proof}

\begin{lemma}
  \label{lemma:trace_quantum}
  \begin{align}
    \sigma_j'^T
    e^{- \frac{\beta}{m} \cH_\text{q}}
    \sigma_{j+1}
    \propto
    e^{s(\sigma_j',\sigma_{j+1}) f(\beta,\Gamma)}
    .
  \end{align}
\end{lemma}
\begin{proof}
  
Substituting $(A\otimes B)(C \otimes D) = (AC) \otimes (BD)$
and $e^{A_1 + A_2} = e^{A_1}e^{A_2}$ when $A_1A_2 = A_2A_1$, we find,
\begin{align}
  \sigma_j'^T
  e^{- \frac{\beta}{m} \cH_\text{q}}
  \sigma_{j+1}
  =&
  \sigma_j'^T
  \left(
  \bigotimes_{i=1}^n 
  e^{- \frac{\beta}{m} \rho_i}
  \right)
  \sigma_{j+1}
  \nn\\
  =&
  \prod_{i=1}^n 
  \tilde{\sigma}_{j,i}'^T
  e^{- \frac{\beta}{m} \rho}
  \tilde{\sigma}_{j+1,i}
  \label{eq:trace_quantum_after_Minka}
  ,
\end{align}
where $\tilde{\sigma}_{j,i}$ is the $i$-th element of Kronecker
product of $\sigma_j$, s.t. $\sigma_j = \bigotimes_{i=1}^n
\tilde{\sigma}_{j,i}$.  Substituting the following
\eqref{eq:trace_quantum_support1} and
\eqref{eq:trace_quantum_support2} into
\eqref{eq:trace_quantum_after_Minka},
\begin{align}
  e^{- \frac{\beta}{m} \rho}
  =&
  \sum_{l=0}^\infty \frac{1}{l!}
  \left( - \frac{\beta}{m} \right)^l \rho^l
  \label{eq:trace_quantum_support1}
  \\
  \rho^l =& \Gamma^l (\bbE_k - \bbone_k)^l
  = \Gamma^l\left\{ \sum_{i=0}^l \binomialCoefficient{l}{i} \bbE_k^i (- \bbone_k)^{l-i}
  \right\}
  \nn\\
  =& \Gamma^l\left\{ \bbE_k + \sum_{i=0}^{l-1} \binomialCoefficient{l}{i} k^{l-i-1} (-1)^{l-i} \bbone_k
  \right\}
  \nn\\
  =& \Gamma^l\left\{ \bbE_k + \frac{1}{k}\sum_{i=0}^{l-1} \binomialCoefficient{l}{i} (-k)^{l-i} \bbone_k
  \right\}
  \nn\\
  =& \Gamma^l\left\{ \bbE_k + \frac{1}{k}\left\{ (1-k)^l -1 \right\} \bbone_k 
  \right\}
  ,
  \label{eq:trace_quantum_support2}
\end{align}
we have
\begin{align}
  &
  \sigma_j'^T
  e^{- \frac{\beta}{m} \cH_\text{q}}
  \sigma_{j+1}
  \nn\\
  &=
  \prod_{i=1}^n 
  \sum_{l=0}^\infty \frac{1}{l!} \! \left( \! -\frac{\beta\Gamma}{m}\right)^l
  \!
  \tilde{\sigma}_{j,i}'^T
  \!
  \left( \bbE_k + \frac{1}{k}\left\{ (1-k)^l -1 \right\} \bbone_k  \! \right)
  \!
  \tilde{\sigma}_{j+1,i}
  \nn\\
  &=
  \prod_{i=1}^n 
  \sum_{l=0}^\infty \frac{1}{l!} \! \left(\!-\frac{\beta\Gamma}{m}\right)^l
  \!\!
  \left\{
  \delta(\tilde{\sigma}_{j,i}', \tilde{\sigma}_{j+1,i})
  +
  \frac{1}{k}(1-k)^l - \frac{1}{k}
  \right\}
  \nn\\
  &=
  \prod_{i=1}^n 
  \left\{
  e^{-\frac{\beta\Gamma}{m}}
  \delta(\tilde{\sigma}_{j,i}', \tilde{\sigma}_{j+1,i})
  +
  \frac{1}{k}e^{-\frac{\beta\Gamma}{m}(1-k)}
  -
  \frac{1}{k}e^{-\frac{\beta\Gamma}{m}}
  \right\}
  \nn\\
 &\propto
  e^{s(\sigma_j',\sigma_{j+1}) f(\beta,\Gamma)}
  .
  \label{eq:trace_quantum}
\end{align}
\end{proof}

Using Lemma \ref{lemma:trace_classical} and Lemma \ref{lemma:trace_quantum} into
\eqref{eq:qa_naive_st_expansion}, \eqref{eq:trotter_expansion_inner} becomes,
\begin{align}
  &
  \frac{1}{Z}
  \sigma_1^T
  \left(
  e^{- \frac{\beta}{m} \cH_\text{c}}
  e^{- \frac{\beta}{m} \cH_\text{q}} 
  \right)^m
  \sigma_1
  \nn\\&
  =
  \frac{1}{Z}
  \sum_{\sigma_2}
  ...
  \sum_{\sigma_m}
  \prod_{j=1}^m
  p_\text{SA}(\sigma_j; \beta/m)
  e^{s(\sigma_j,\sigma_{j+1}) f(\beta,\Gamma)}
  .
  \nn
\end{align}
From \eqref{eq:trotter_expansion} and the above expression, we
have shown Theorem \ref{theorem:qa_is_approximated_by_st}.




{\small
\begin{thebibliography}{11}
\providecommand{\natexlab}[1]{#1}
\providecommand{\url}[1]{\texttt{#1}}
\expandafter\ifx\csname urlstyle\endcsname\relax
  \providecommand{\doi}[1]{doi: #1}\else
  \providecommand{\doi}{doi: \begingroup \urlstyle{rm}\Url}\fi

\bibitem[Blei et~al.(2003)Blei, Ng, and Jordan]{Blei03}
D.~M. Blei, A.~Y. Ng, and M.~I. Jordan.
\newblock Latent {D}irichlet allocation.
\newblock \emph{Journal of Machine Learning Research}, 3:\penalty0 993--1022,
  2003.

\bibitem[Geman and Geman(1984)]{Geman84}
Stuart Geman and Donald Geman.
\newblock Stochastic relaxation, {G}ibbs distributions, and the {B}ayesian
  restoration of images.
\newblock \emph{IEEE Pattern Analysis and Machine Intelligence,}, 6:\penalty0
  721--741, 1984.

\bibitem[Apolloni et~al.(1989)Apolloni, Carvalho, and Falco]{Apolloni89}
B.~Apolloni, C.~Carvalho and D.~de Falco
\newblock Quantum Stochastic Optimization.
\newblock \emph{Stochastic Processes and their Applications}, 33,
		233--244, 1989.

\bibitem[Kadowaki and Nishimori(1998)]{Kadowaki98}
Tadashi Kadowaki and Hidetoshi Nishimori.
\newblock Quantum annealing in the transverse {I}sing model.
\newblock \emph{Physical Review E}, 58, 5355, 1998.

\bibitem[Kirkpatrick et~al.(1983)Kirkpatrick, Gelatt, and
  Vecchi]{Kirkpatrick83}
S.~Kirkpatrick, C.~D. Gelatt, and M.~P. Vecchi.
\newblock Optimization by simulated annealing.
\newblock \emph{Science}, 220\penalty0 (4598):\penalty0 671--680, 1983.

\bibitem[Liang et~al.(2007)Liang, Jordan, and Taskar]{Liang07}
Percy Liang, Michael~I. Jordan, and Ben Taskar.
\newblock A permutation-augmented sampler for {DP} mixture models.
\newblock In \emph{ICML}, pages 545--552. Omnipress, 2007.

\bibitem[Matsuda et~al.(2009)Matsuda, Nishimori, and Katzgraber]{matsuda08}
Yoshiki Matsuda, Hidetoshi Nishimori, and Helmut~G Katzgraber.
\newblock Ground-state statistics from annealing algorithms: Quantum vs
  classical approaches.
\newblock arXiv:0808.0365, 2009.

\bibitem[Minka(2000)]{Minka00}
Thomas~P. Minka.
\newblock Old and new matrix algebra useful for statistics, 2000.

\bibitem[Richardson and Green(1997)]{Richardson97}
Sylvia Richardson and Peter~J. Green.
\newblock On {B}ayesian analysis of mixtures with an unknown number of
  components.
\newblock \emph{Journal of the Royal Statistical Society, Series B},
  59\penalty0 (4):\penalty0 731--792, 1997.

\bibitem[Santoro et~al.(2002)Santoro, Marto\v{n}\'{a}k, Tosatti, and
  Car]{Santoro02}
Giuseppe~E. Santoro, Roman Marto\v{n}\'{a}k, Erio Tosatti, and Roberto Car.
\newblock Theory of quantum annealing of an {I}sing spin glass.
\newblock \emph{Science}, 295\penalty0 (5564):\penalty0 2427--2430, 2002.

\bibitem[Suzuki(1976)]{Suzuki76}
Masuo Suzuki.
\newblock Relationship between d-dimensional quantal spin systems and
  (d+1)-dimensional {I}sing systems -- equivalence, critical exponents and
  systematic approximants of the partition function and spin correlations --.
\newblock \emph{Progress of Theoretical Physics}, 56\penalty0 (5):\penalty0
  1454--1469, 1976.

\bibitem[Trotter(1959)]{Trotter59}
H.~F. Trotter.
\newblock On the product of semi-groups of operators.
\newblock \emph{Proceedings of the American Mathematical Society}, 10\penalty0
  (4):\penalty0 545--551, 1959.

\end{thebibliography}

}
\end{document}